\documentclass[aps,prx,twocolumn,showpacs,psfig,superscriptaddress,longbibliography]{revtex4-1}
\usepackage{listings}
\usepackage{tikz}

\usepackage{amssymb}
\usepackage{algpseudocode}
\usepackage{algorithm}
\usepackage{amsmath}
\usepackage{braket}
\usepackage{booktabs}
\usepackage{chngcntr}
\usepackage{color}
\usepackage{comment}
\usepackage{dsfont}
\usepackage{etoolbox}
\usepackage{epstopdf}
\usepackage[T1]{fontenc}
\usepackage{float}
\usepackage{graphicx}
\usepackage[colorlinks=true,linkcolor=blue,citecolor=blue,urlcolor=blue]{hyperref}
\usepackage[latin9]{inputenc}
\usepackage{indentfirst}
\usepackage{latexsym,bm,euscript}
\usepackage{mathrsfs}
\usepackage{multirow}
\usepackage{natbib}
\usepackage{soul}
\usepackage{subfigure}
\usepackage{times}
\usepackage{titlesec}
\usepackage{tikz}
\usepackage{txfonts}
\usepackage{xspace}
\usepackage{xfrac}

\usepackage{mathtools}
\usepackage[vlined, ruled, algo2e, linesnumbered]{algorithm2e}

\def\beqn{\begin{eqnarray}}
\def\eeqn{\end{eqnarray}}
\def\beq{\begin{equation}}
\def\eeq{\end{equation}}

\newcommand{\Eq}[1]{Eq.~\eqref{#1}}

\newcommand{\Fig}[1]{Fig.~\ref{#1}}

\newcommand*{\tran}{^{\mkern-1.5mu\mathsf{T}}}

\graphicspath{{./Fig/}{./Fig/tst_L06/}{./Fig/tst_L24/}}

\begin{document}
\title{Learning Effective Spin Hamiltonian of Quantum Magnet}

\author{Sizhuo Yu}
\thanks{These authors contributed equally to this work.}
\affiliation{School of Physics and Key Laboratory of Micro-Nano Measurement-Manipulation and Physics (Ministry of Education), 
Beihang University, Beijing 100191, China}

\author{Yuan Gao}
\thanks{These authors contributed equally to this work.}
\affiliation{School of Physics and Key Laboratory of Micro-Nano Measurement-Manipulation and Physics (Ministry of Education), 
Beihang University, Beijing 100191, China}

\author{Bin-Bin Chen}
\affiliation{School of Physics and Key Laboratory of Micro-Nano Measurement-Manipulation and Physics (Ministry of Education), 
Beihang University, Beijing 100191, China}

\author{Wei Li}
\email{w.li@buaa.edu.cn}
\affiliation{School of Physics and Key Laboratory of Micro-Nano Measurement-Manipulation
and Physics (Ministry of Education), Beihang University, Beijing 100191, China}
\affiliation{International Research Institute of Multidisciplinary Science, Beihang University, Beijing 100191, China}

\begin{abstract} 
Interacting spins in quantum magnet can cooperate and 
exhibit exotic states like the quantum spin liquid. 
To explore the materialization of such intriguing states, 
the determination of effective spin Hamiltonian of the 
quantum magnet is thus an important, while at the same time, 
very challenging inverse many-body problem. 
To efficiently learn the microscopic spin Hamiltonian 
from the macroscopic experimental measurements, 
here we propose an unbiased Hamiltonian searching 
approach that combines various optimization strategies, 
including the automatic 
differentiation and Bayesian optimization, etc, 
with the exact diagonalization and many-body 
thermal tensor network calculations. 
We showcase the accuracy and powerfulness by 
applying it to training thermal data generated 
from a given spin Hamiltonian, and then to realistic 
experimental data measured in the spin-chain 
compound Copper Nitrate and triangular-lattice 
materials TmMgGaO$_4$. This automatic 
Hamiltonian searching constitutes a very 
promising approach in the studies of 
the intriguing spin liquid candidate 
magnets and correlated electron materials in general. 
\end{abstract}

\date{\today}
\maketitle

\textit{Introduction.---}
Exotic many-body quantum states and phenomena 
in magnetic materials have raised great research interest recently. 
Among others, an intriguing topic is the materialization of  
quantum spin liquids with topologically ordered 
ground states and anyonic excitations, which
has been long pursued in quantum magnetism
~\cite{Anderson1973,Kitaev2006,Zhou2017,Balents2010}. 
Some prominent spin liquid candidate 
systems include the kagome~\cite{Han2012,Fu2015}, 
triangular~\cite{Shimizu2003,Yamashita2010,Liu2018}, 
and Kitaev magnets~\cite{Jackeli2009,Chaloupka2010,Ye2012,
Banerjee2017}. However, the lack of precise knowledge 
on the effective spin lattice models of these 
frustrated magnets hinders the unambiguous 
understanding of the quantum states and phases therein. 

The identification of the microscopic spin model and 
the determination of Hamiltonian parameters of 
the magnetic materials constitute an important step 
towards understanding their properties. It is, however, 
a very challenging problem to ``learn'' the spin 
Hamiltonian from experimental measurements.
For example, to understand the quantum states 
in the prominent Kitaev materials $\alpha$-RuCl$_3$,  
various spin models have been proposed, yet none
of them could satisfactorily explain all experimental 
observation~\cite{Laurell2020}. 
The difficulty is two-fold. 
Firstly, to solve the spin Hamiltonian
and compute the thermodynamic and dynamic properties
that are experimentally relevant is by no means an easy 
problem, as there is a many-body exponential wall to break.
Secondly, even worse, the determination of the effective spin 
Hamiltonian from experimental measurements 
constitutes an inverse many-body problem. 

\begin{figure}[!htbp]
\includegraphics[angle=0,width=1\linewidth]{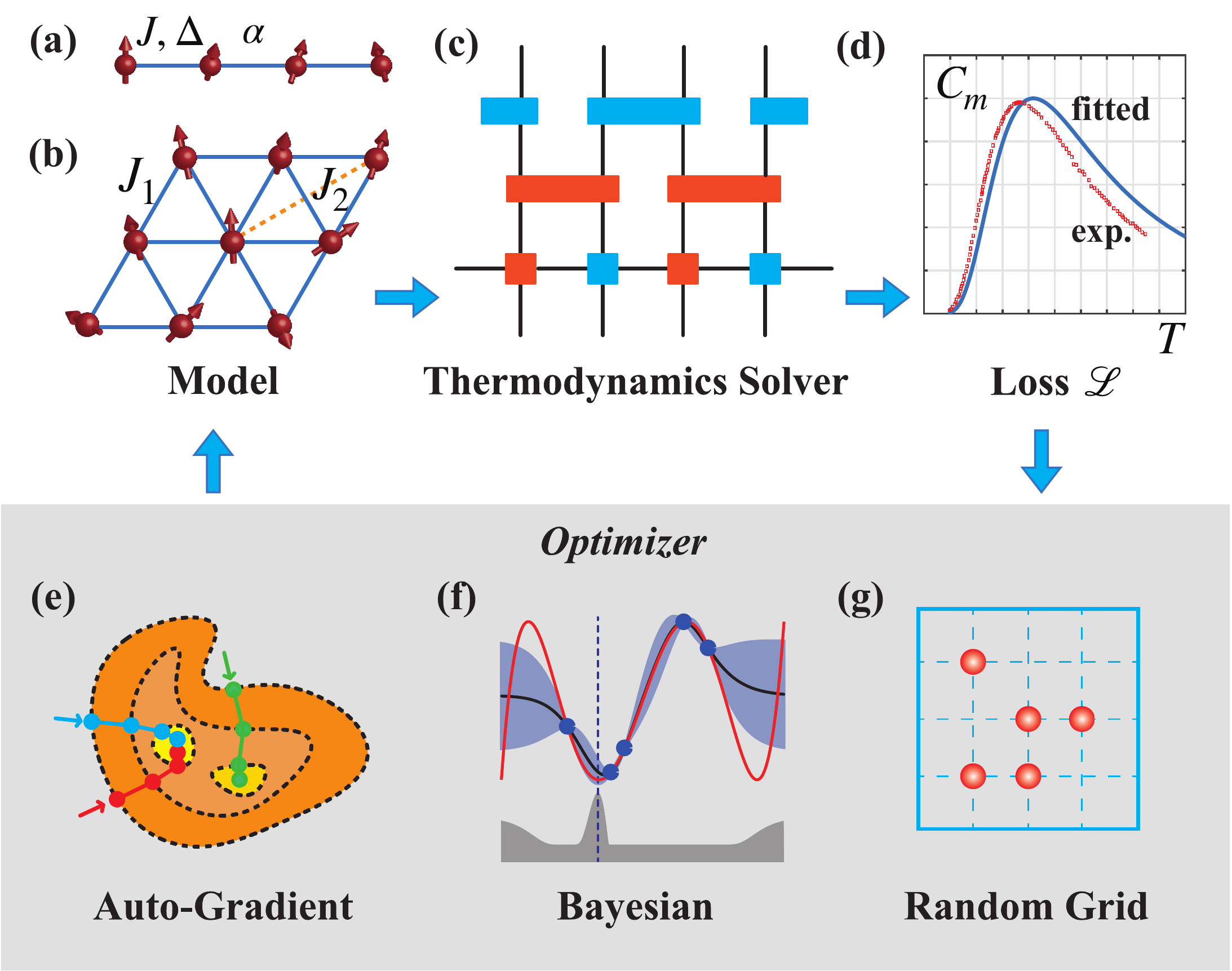}
\caption{The workflow of automatic 
Hamiltonian searching. (a,b) show the HAFC model with coupling $J$, 
ratio $\alpha$ and magnetic anisotropy $\Delta$, 
and frustrated triangular-lattice model with nearest-neighboring 
coupling $J_1$ and next-nearest-neighboring coupling $J_2$. 
With the thermodynamics solvers in (c), we can compute the 
(d) loss function $\mathcal{L}$, and feed it to the 
optimizers including (e) auto-gradient, 
(f) Bayesian, and (g) random grid methods.
The optimizer proposes a new trial parameter set
for the next iteration, until the convergence 
criteria is reached.
}
\label{Fig:Illustr}
\end{figure}

The recent progress in finite-temperature tensor networks 
has been swift, which enables efficient and accurate 
calculations of the thermodynamic properties of large-scale 
1D and 2D systems down to low temperature~\cite{Bursill1996,
Wang1997,Xiang1998,Feiguin2005,White2009,Stoudemire2010,
Li2011,Dong2017,Chen2017,Chen2018,Lih2019}. 
Nevertheless, these thermal tensor network calculations
generically demands considerable computational resources
for low-temperature simulations. Therefore, considering
a realistic magnetic material [c.f. Eqs.~(\ref{Eq:HAFC},
\ref{Eq:CN}, \ref{Eq:TMGO}) below], grid searching 
by computing the many-body systems point by point 
in the parameter space and compare to to experimental data, 
is a very laborious and, even unfeasible for Hamiltonians 
with, say, more than 5 parameters in practice.

Machine learning techniques have recently brought into 
quantum many-body computations very helpful new 
perspectives and methodology. For example, 
it has been proposed that the artificial neural 
networks can serve as a powerful variational 
many-body wavefunction ansatz that
produces accurate results~\cite{Carleo2017}, 
and the differentiable tensor network approach 
helps to design novel tensor renormalization 
group algorithms with improvement~\cite{Liao2019,Chen2020}. 
On the other hand, the many-body tensor network 
approaches have also found their applications in
machine learning, including the matrix product state
and tree tensor network based supervised learning~\cite{Stoudenmire2016,Liu2019}, 
the Bayesian tensor-network probabilistic learning
~\cite{Ran2020}, and many others~\cite{Cichocki2017,Han2018,Glasser2019ExpressivePO}.

In this work, we propose an automatic Hamiltonian searching approach 
for determining the effective spin model --- the magnetism genome ---
from fitting thermodynamic data of quantum magnetic materials.
Our method explores the parameter space efficiently, 
with gradient optimization by automatic differentiation 
(auto-gradient) and Bayesian optimization schemes, 
inspired by machine learning techniques. In particular,
the predicted landscape of loss function in the 
parameter space can present a comprehensive 
information, and is thus of great helpfulness in, e.g.,
reducing the human bias in the parameter fittings. 
The automatic Hamiltonian searching, 
given it auto-gradient or Bayesian, 
are very flexible and can be combined 
with various many-body methods, 
ranging from small-size exact diagonalization 
(ED, as a high-$T$ solver) to large-scale 
(even infinite-size) thermal tensor networks 
(low-$T$ solver)~\cite{Li2011,Dong2017,Chen2018,Lih2019},
and other thermodynamics solvers \cite{SM}.

\textit{Thermodynamics many-body solver.---} 
When only high-$T$ thermal data are involved, 
the ED calculations can be employed to compute 
the spin lattice model with limited system sizes. 
The effective thermal correlation length is short,
and it thus serves only as a high-$T$ solver. 
Nevertheless, we find ED calculations are already very
helpful for automatic determination of the spin Hamiltonians,
as the valuable correlations and thus interactions information 
``hidden'' in the quantitative details of the thermodynamic curves 
(though featureless to human eyes) can be efficiently 
extracted by optimization techniques widely used in machine learning.

Moreover, to unambiguously determine the spin Hamiltonian,
we employ large-scale tensor network methods
as the low-$T$ thermodynamic solver. Linearized tensor 
renormalization group (LTRG)~\cite{Li2011,Dong2017} can 
compute infinite-length system and thus provide an accurate  
access to the full-temperature range of spin-chain materials. 
Beyond 1D system, other thermal tensor network methods 
including the exponential tensor renormalization group~\cite{Chen2018,Lih2019},
and tensor product state approaches~\cite{Li2011,Czarnik2014}
can be used to compute large-scale 2D systems, 
which can also be conveniently combined with either 
auto-gradient or Bayesian optimization schemes will 
be discussed below shortly.

\begin{figure}[!htbp]
\includegraphics[angle=0,width=1\linewidth]{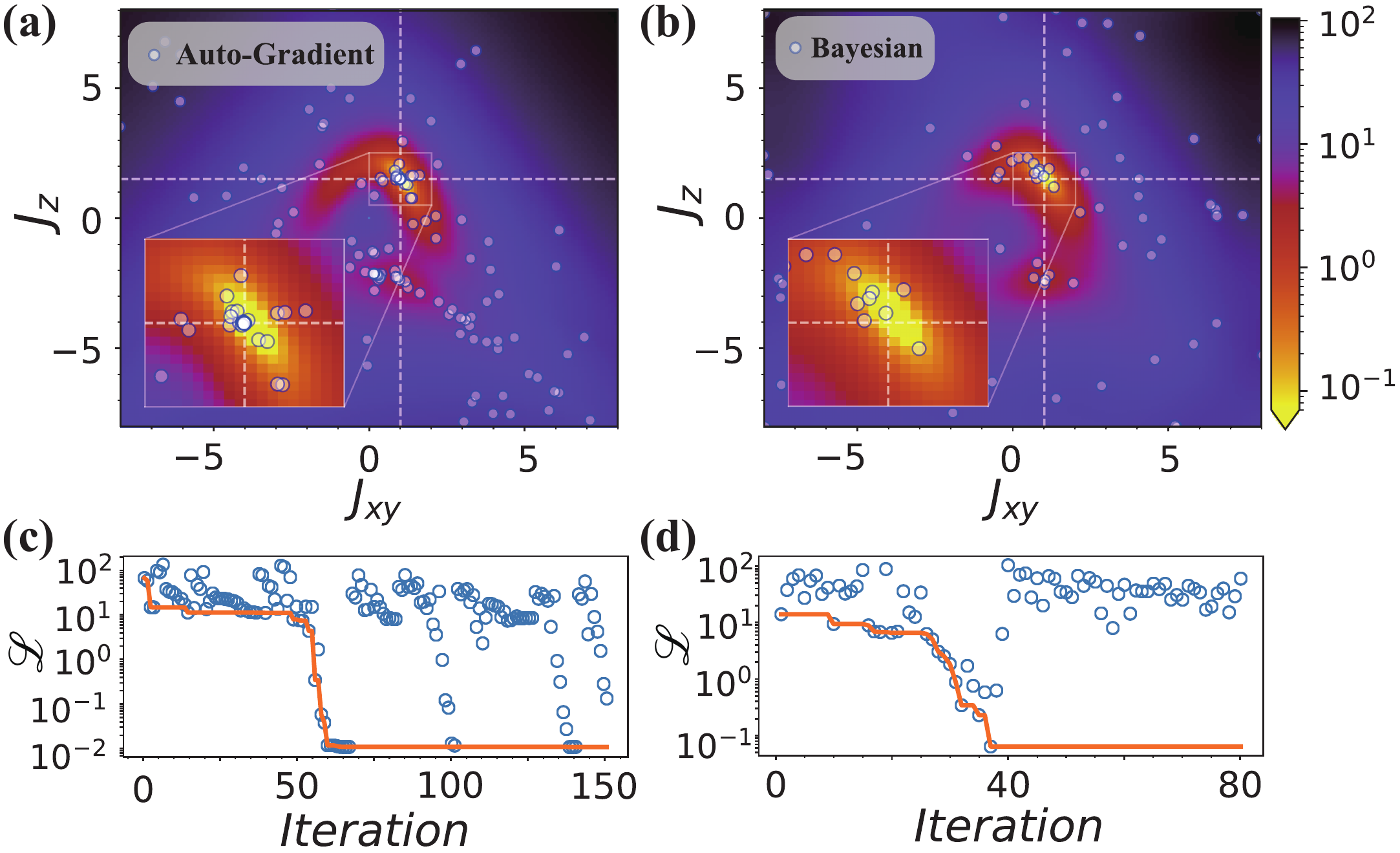}
\caption{(a)  
The scatters indicate the evaluated queries of 150 iterations of multi-restart 
gradient optimization processes, and the background, $\mathcal{L}$ landscape, 
is obtained via a grid search. (b) Landscape of $\mathcal{L}$ predicted by the 
Bayesian optimization with an evaluated query of 80 parameter points, where 
the loss along indicated dash lines can be found in Supplementary Fig.~S4 (b,c). 
(c, d) Solid lines indicate the convergence of $\mathcal{L}$ using respectively 
multi-restart auto-gradient and Bayesian optimizations, and the scatters represent 
the evaluated function value $\mathcal{L}(\mathbf{x}_i)$ at each iteration.}
\label{Fig:Land}
\end{figure}

\begin{figure}[!htbp]
\includegraphics[angle=0,width=1\linewidth]{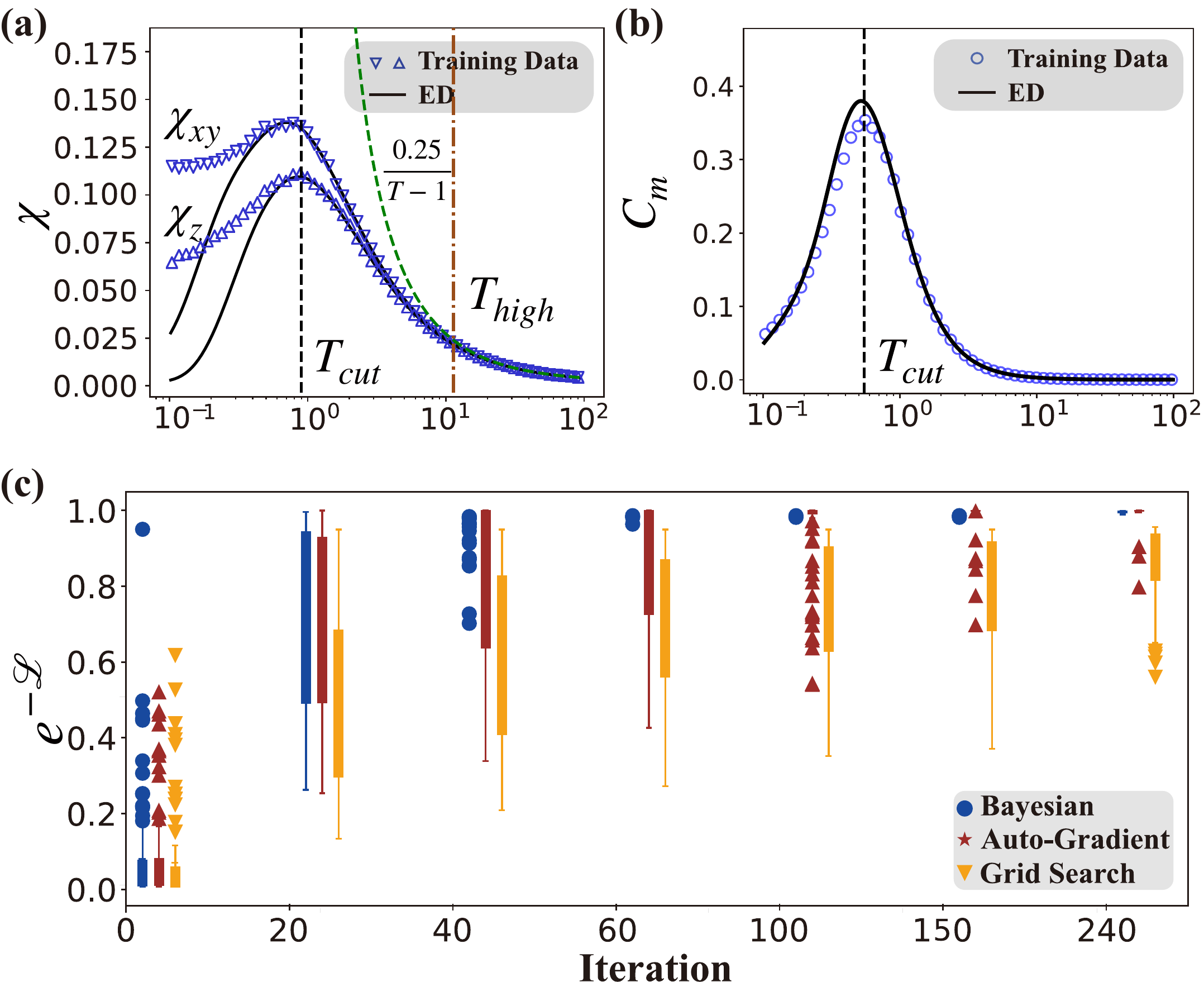}
\caption{(a) The in-plane $\chi_{xy}^{\,}$ 
and out-of-plane $\chi_{z}^{\,}$ of the 
training data generated by an infinitely-long 
XXZ chain with $J_z  = 1.5, J_{xy} = 1$ 
(hollow symbols) and a best fitting based 
on 10-site ED calculations with parameters 
$J_z = 1.49(1), J_{xy}= 1.02(1)$ (solid line).
Only ``experimental" data with temperatures 
higher than $T_{\rm cut}$ (black dashed line) 
are involved in the fittings. Below the 
temperature scale $T_{high}$ (brown dash-dotted line), 
the susceptibility $\chi$ deviates from the Curie-Weiss 
behaviors marked with the green dashed line. 
(b) The magnetic specific heat $C_m$ of training 
experiment and optimal ED fitting. (c) The box plot of 
best $\mathcal{L}$ found at $n$-th iteration 
of 100 independent experiments 
with three optimization schemes.}
\label{Fig:HAFC}
\end{figure}

\begin{figure}[htbp]
\includegraphics[angle=0,width=1\linewidth]{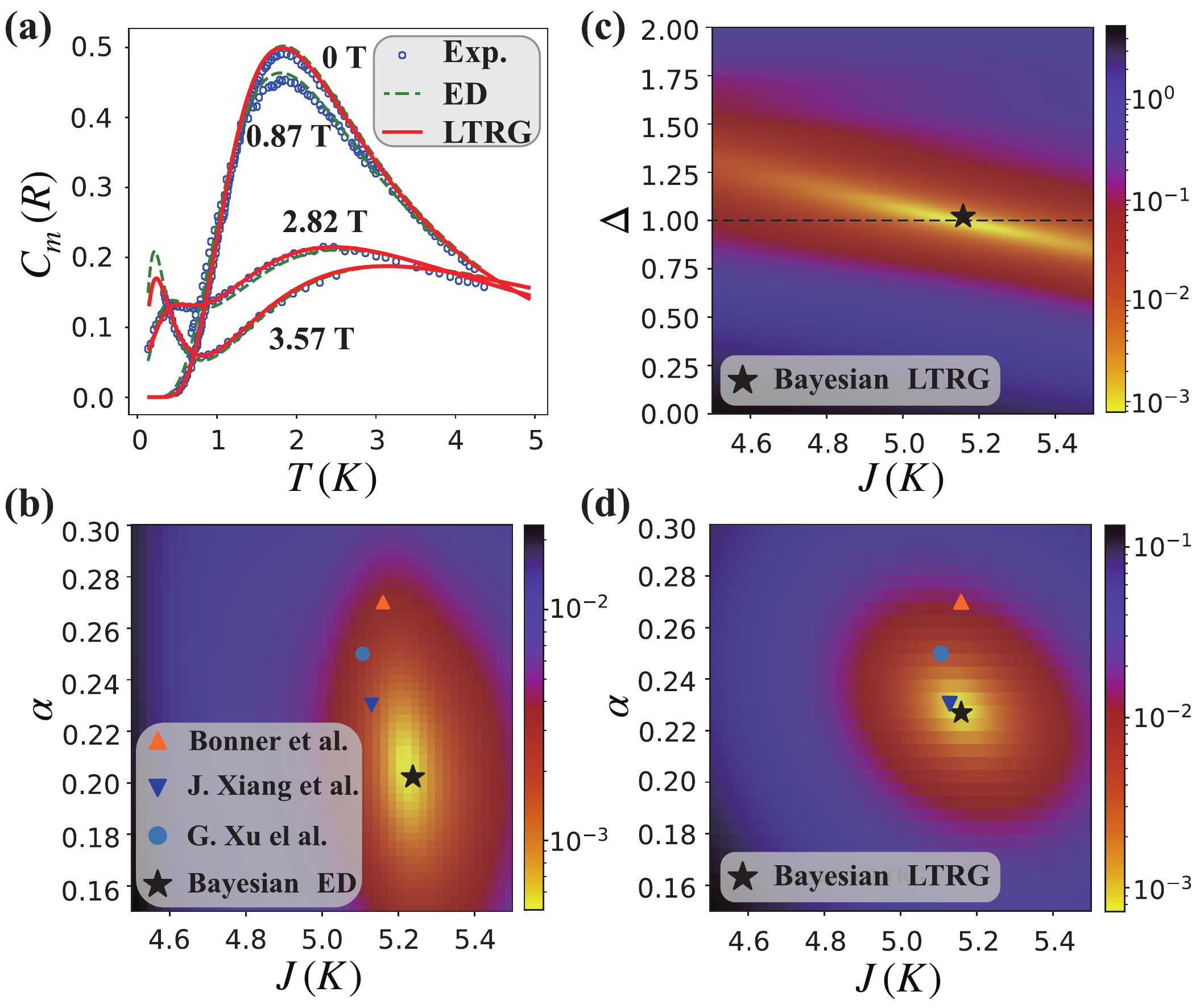}
\caption{(a) Magnetic specific heat $C_m /R$ at various fields of 0, 0.87, 2.82, and 3.57 T. 
The dashed lines represent the ED fittings and the solid lines are LTRG calculations. 
(b) shows the Bayesian $\mathcal{L}$ landscape within the $J-\alpha$ plane, using ED solver, 
where the estimated optimal parameter point (the asterisk) are compared to results in 
previous studies~\cite{vanTol1971,Xu2000,Xiang2017}. 
(c, d) The Bayesian $J-\alpha$ (with fixed $\Delta = 1$, $g=2.31$) and 
$J-\Delta$ ($\alpha = 0.23$, $g=2.31$ fixed) landscape, obtained after 
400 iterations of LTRG calculations. The optimal parameter found is $J = 5.16(2)$ K, 
$\alpha = 0.227(3)$, $\Delta = 1.01(1)$, $g=2.237(8)$, which are very 
close to the estimated parameters in Ref.~\cite{Xiang2017}, 
and has a slightly smaller loss $\mathcal{L} = 7.4 \times 10^{-4}$.}
\label{Fig:CN}
\end{figure}

\begin{figure*}[!htb]
\includegraphics[angle=0,width=1\linewidth]{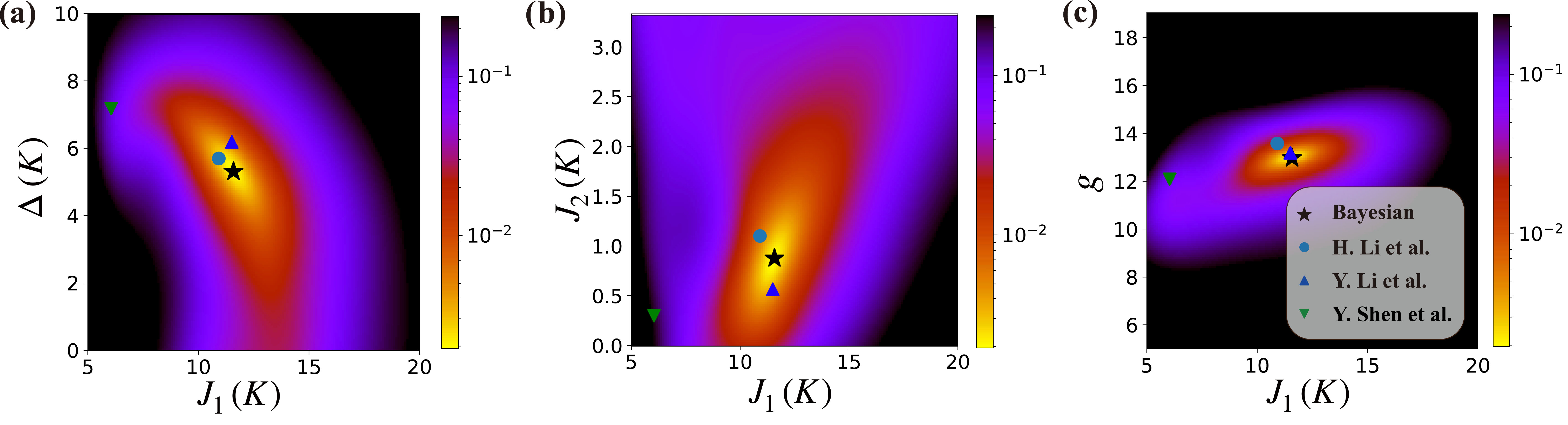}
\caption{Loss landscape in
(a) $J_1$-$\Delta$, (b) $J_1$-$J_2$, and 
(c) $J_1$-$g$ planes, obtained after 300 
iterations of Bayesian optimization.
The estimated Hamiltonian parameters
with $J_1=11.5(1)$~K, $J_2=0.89(7)$~K, $\Delta=5.32(6)$~K, 
and $g=13.00(3)$, are labeled by the asterisk and
compared to the solid circle, upper, and lower triangles 
that represent the previously fitted parameters 
from Refs.~\cite{Lih2020,Li2020,Shen2019}. 
}
\label{Fig:TMGO}
\end{figure*}

\textit{Random grid, auto-gradient and Bayesian optimization.---}
The objective loss function of the thermal data fitting reads
\begin{equation}
\mathcal{L}(\mathbf{x}_i) = \sum_{\alpha}\frac{1}{N_\alpha} \lambda_{\alpha}(\frac{O^{\rm exp}_\alpha-O^{\rm sim}_\alpha}{O^{\rm sim}_\alpha})^2,
\label{Eq:Loss}
\end{equation}
where $O^{\rm exp}_\alpha$ and $O^{\rm sim}_\alpha$ 
(with $\alpha$ labeling different physical quantities)
are the experimental and simulated quantities, respectively,
and $\lambda_{\alpha}$ is an empirical weight coefficient 
set to unity by default. The parameter vector 
$\mathbf{x}$ contains various components 
including $J$, $\Delta$ and $g$, and
span a parameter space $\mathcal{X}$.
$N_\alpha$ is the data point number of quantity 
$O_\alpha$, and thus $1/N_\alpha$ normalizes 
the loss function per point 
\footnote{In practice, we first employed a loss function without the 
denominator $1/ O^{\rm sim}_\alpha$ in Figs.~\ref{Fig:Land}
and~\ref{Fig:HAFC}, and then follows the exact form as
Eq.~(\ref{Eq:Loss}) in the cases of Figs.~\ref{Fig:CN} and
\ref{Fig:TMGO}. Both schemes work well, and the 
design of the loss function have an empirical impact 
on its overall shape over the parameter 
space $\mathcal{X}$, whose effects in the optimization 
efficiency will be carefully addressed in future studies.}.

An efficient optimizer that minimizes 
the loss function $\mathcal{L}$
in the parameter space $\mathcal{X}$ plays an 
indispensable role in the automatic Hamiltonian searching. 
In this work, we have employed two machine-learning
inspired algorithms: auto-gradient and Bayesian searching, 
and compare them to a plain random grid method \cite{SM}.

In particular, inspired by the backpropagation 
arithmetic in deep learning~\cite{LeCun2015}, 
automatic differentiation has been introduced 
into tensor-network methods for quantum many-body 
computations~\cite{Liao2019, Chen2020}. Here in our work,
to obtain the gradient information that greatly facilitates 
the search of spin Hamiltonians, we realize the 
differentiable programming of the thermodynamics solver. 
The basic idea is that, given the many-body solver fully 
differentiable, the derivatives between intermediate variables 
of adjacent steps are stored in the forward process all the 
way to the final loss function $\mathcal{L}$. 
Given that, the derivatives of the loss function respective 
to the Hamiltonian parameters, ${\bf{\bar x}_i} = 
\partial \mathcal{L}/\partial{\bf{x}_i}$, can be 
computed automatically following the derivative chain 
rule in the backward propagations, which can be further 
utilized to optimize the parameters $\bf{x}_i$ 
via gradient-based optimizer \cite{SM}.
As the loss $\mathcal{L}$ 
is generically non-convex (c.f. Fig.~\ref{Fig:Land}), 
we need to restart and perform the auto-gradient 
search for multiple times, in order to guarantee 
the convergence to global minimum.

The Bayesian optimization (BO) is a powerful and highly 
efficient method which has been widely used in 
hyper-parameter tuning of deep neural networks, 
active, and reinforce learning, etc~\cite{7352306}. 
As most of the state-of-the-art thermodynamics 
many-body solvers are computationally costly, 
it is then essential to exploit the information of tested 
parameter points and determine where to evaluate 
the function next~\cite{Melnikov1221}. 

In practice, BO minimizes our loss function $\mathcal{L}$ 
by iteratively updating a statistical model 
$\mathcal{GP} : \mathcal{X} \rightarrow \mu, \sigma$ over the entire 
parameter space $\mathcal{X}$, and $\mu, \sigma$ 
represent the predicted value and uncertainty, 
as shown in Fig.~\ref{Fig:Illustr}(f). 
The parameters $\bf{x}$ to be evaluated 
at each iteration is determined by maximizing 
an acquisition function $\alpha_{\rm EI}(\mathbf{x})$, 
based on the expected improvement. 
To be specific, one can determine 
$\mathbf{x}_{n+1} = \arg \max \alpha_{\rm EI}(\mathbf{x}) 
= \arg \max \mathbb{E}[\mathcal{L}_{n,min} - \mu_n(\mathbf{x})]$ 
as the best parameter candidate in the next ($n+1$) iteration,
where $\mathcal{L}_{n,min}$ denotes the 
minimal loss function found in the $n$-th iteration. 
This method can elegantly balance the optimization efficiency 
and the exploration of parameter space $\mathcal{X}$ by 
choosing the appropriate acquisition criteria \cite{SM}.

\textit{Refind the spin Hamiltonian.---}
We start with training thermal data generated from the XXZ 
Heisenberg antiferromagnetic chain (HAFC) model with 
a given parameter, and feed the ``experimental'' data to 
various optimizers, i.e., random grid, auto-gradient, 
and the Bayesian searching, to see if we can refind the 
correct Hamiltonian parameters. Below, we stick to an 
thermodynamics ED solver, and focus on the comparison 
between various optimization schemes.

The thermodynamic data of HAFC systems are computed 
from the model Hamiltonian below, i.e.,
\begin{equation}
\label{Eq:HAFC}
H = \sum_{\langle i,j \rangle} J_{xy} (S_i^{x} S_j^{x} + S_i^y S_j^y) 
+ J_z S_i^z S_j^z,
\end{equation}
where $\langle i, j \rangle$ represents a nearest-neighboring 
pair of sites. We employ LTRG to generate the infinite-chain
thermal data of HAFC with $J_{xy}=1$ and $J_z =1.5$ 
(for cases with different $J_z$ values, 
see Supplementary Fig.~S4).
Gaussian noises $\mathcal{N} (0, 0.01 \times E_i)$ 
are added to each data point of mean value $E_i$
are also introduced (c.f. Fig.~\ref{Fig:HAFC}), 
to mimic the measurement errors in real experiments.
We show below that the smart optimizers 
and the high-$T$ ED solver can cooperate 
and do a surprisingly good job to 
``learn'' the correct Hamiltonian parameters.

As shown in Fig.~\ref{Fig:Land}(a), 
the loss function landscape scanned throughout
the whole parameter space $\mathcal{X}$
is found to have a global minimal at around 
$J_{xy}=1$ and $J_z=1.5$, 
exactly the input model parameter set, 
which delivers a key information that one can, 
in principle, locate the correct interaction parameters
even from high-$T$ thermodynamics. 
Indeed, both the auto-gradient and 
BO schemes can efficiently 
and accurately find the original parameters. 
The latter can also reproduce the correct 
loss landscape, c.f. Fig.~\ref{Fig:Land}(a,b). 
In the automatic Hamiltonian searching, 
as the ED thermodynamics solver can only 
simulate relatively high-$T$ properties, so
we introduce a cut-off temperature $T_{\rm cut}$
in the fitting. As shown in Fig.~\ref{Fig:HAFC}(a,b),
we only fit thermal data at $T \gtrsim T_{\rm cut} \simeq O(1)$,
which are chosen as the peak positions of magnetic 
susceptibility and specific heat curves, respectively.
The dependence of determined Hamiltonian parameters
on $T_{\rm cut}$ is discussed in the Supplementary\cite{SM}.

Notably, in the definition of $\mathcal{L}$, 
c.f. Eq.~(\ref{Eq:Loss}), when only 
$C_m$ and $\chi_{z}$ are included, 
the optimizers can find two 
optimal parameters $J_{xy}= \pm 1$ and $J_z=1.5$, 
which is very interesting as indeed the two parameter points
have exactly the same thermodynamic traits,
as the Hamiltonian Eq.~(\ref{Eq:HAFC})
has the same energy spectra for $J_{xy}= \pm 1$,
and the our smart approach can automatically find 
this fact out. Nevertheless, higher resolution can 
be achieved by adding more thermal data to the fittings.
The two-fold degeneracy in landscape can be
removed once $\chi_{xy}$ is introduced to $\mathcal{L}$.
As a result, in Fig.~\ref{Fig:Land}(a,b) 
and Fig.~\ref{Fig:HAFC} we have included 
the specific heat $C_m$, both in-plane and 
out-of-plane magnetic susceptibilities 
$\chi_{xy}$ and $\chi_{z}$, 
and the model parameters is now uniquely 
pinpointed \cite{SM}.

From Fig.~\ref{Fig:HAFC}(c), 
in the 100 independent searching experiments,
we note that both the Bayesian and 
auto-gradient approaches clearly outperforms 
the random grid method in both efficiency and 
accuracy [c.f. also Fig.~\ref{Fig:Land}(c,d)].
Although the auto-gradient method can lead to 
very accurate estimate in the ``lucky" case 
(c.f. Fig.~\ref{Fig:Land}), it also has good chance
to be trapped in the local minimal, especially when
the optimization iteration number is relatively small. 
On the other hand, the Bayesian optimization 
is mostly stable amongst three schemes, 
and it finds the optimal parameters
$J_{xy}= 1.025(9)$ and $J_z = 1.49(1)$ very efficiently. 
Due to this reason, and also that the Bayesian
optimization is more flexible and can be combined
with various many-body solvers, below we mainly 
adopt the Bayesian approach and apply it to study
realistic magnetic materials.

\textit{Quantum spin-chain material Copper Nitrate.}--- 
Given the successful benchmark calculations on the
training data set, we now move on to a realistic 
spin-chain material Copper Nitrate, 
Cu(NO$_3$)$_2\,\cdot\,$2.5H$_2$O, 
whose magnetic interactions are described by
the alternating Heisenberg XXZ model~[c.f. Fig.~\ref{Fig:Illustr}(a)]
~\cite{Berger1963,vanTol1971,Xu2000,Xiang2017}, i.e.,
\begin{eqnarray}
\label{Eq:CN}
H & = &J \sum_{n=1}^{L/2}[(S_{2n-1}^x S_{2n}^x +S_{2n-1}^y S_{2n}^y  
+ \Delta S_{2n-1}^z S_{2n}^z) \\ \notag
&+ & \alpha \, (S_{2n}^x S_{2n+1}^x +S_{2n}^y S_{2n+1}^y 
+ \Delta S_{2n}^zS_{2n+1}^z)]  \\  \notag
& - & g \mu_B B  \sum_{i=1}^{L} S_i^z.
\end{eqnarray}
Therefore, the problem is to search for the minimal loss 
$\mathcal{L}$ within a four-dimensional parameter space, 
spanned by the parameter vectors $\bf{x}_i$ containing the 
coupling $J$, ratio $\alpha$, magnetic anisotropy $\Delta$, 
and the Land\'e factor $g$. 

In Fig.~\ref{Fig:CN}, we have employed ED and LTRG
as our high- and low-$T$ thermodynamics solver, 
and find the model parameters automatically 
by fitting the specific heat and magnetic susceptibility measurements, 
above the intermediate temperature $T_{\rm cut}$. 
With the ED solver, we find the so-obtained $J-\alpha$ landscape 
[c.f. Fig.~\ref{Fig:CN}(c)] has a relatively narrow 
distribution in $J$ while a large uncertainty in 
alternating ratio $\alpha$. However, by using the 
LTRG thermodynamics solver of infinite chains, 
we get a significantly improved resolution, 
and find the optimal parameters very close to the previously 
determined model parameters by manual fittings~\cite{Xiang2017}. 

In plotting the landscape in Fig.~\ref{Fig:CN}(b,d), 
we fix $\Delta=1$ [or very close to 1 in Fig.~\ref{Fig:CN}(d)], 
as it has been generally believed that the CN constitutes 
an isotropic Heisenberg spin chain \cite{vanTol1971,Xu2000}
(although it has not been carefully examined before).
With the automatic parameter searching, we show in Fig.~\ref{Fig:CN}(c) 
that $\Delta$ lies within a very narrow regime around 1,
and no essential XXZ anisotropy is there in Copper Nitrate. 

\textit{Triangular-lattice quantum Ising magnet TmMgGaO$_4$}.--- Now we switch to a 2D frustrated 
quantum magnet, and take the triangular-lattice 
rare-earth magnet TmMgGaO$_4$ as an example~\cite{Cava2018,Shen2019,Li2020,Lih2020}. 
The precise determination of the spin Hamiltonian 
plays an indispensable role for understanding the 
emergent U(1) symmetry and topological 
Berezinskii-Kosterlitz-Thouless phase transitions 
in this quantum magnet~\cite{Li2020,Hu2020}. 
In previous studies, the effective low-energy 
spin Hamiltonian of TmMgGaO$_4$ 
is found to fall into a triangular-lattice Ising model 
[c.f. Fig.~\ref{Fig:Illustr}(b)], i.e.,
\begin{equation}
H = J_1 \sum_{\langle i,j \rangle}S_i^z S_j^z + J_2 \sum_{\langle\langle i, j'\rangle\rangle} 
S_i^z S_{j'}^z - \Delta \sum_i S_i^x - g \mu_B B \sum_i S_i^z,
\label{Eq:TMGO}
\end{equation}
where $J_1$ and $J_2$ are nearest-neighboring and next-nearest-neighboring
Ising couplings, respectively, $\Delta$ is the intrinsic transverse field 
in the material (due to fine crystal-field splitting), and $g$ is the Land\'e factor.

We explore the $\mathcal{L}$-landscape in Fig.~\ref{Fig:TMGO},
employing a high-$T$ thermodynamics ED solver on a very small 
9-site system (c.f. Supplementary \cite{SM} for more details).
Clearly, we see an optimal parameter point (asterisk) in 
Fig.~\ref{Fig:TMGO}, which is in very good consistent 
with two of previous model parameter estimates~\cite{Li2020,Lih2020}, 
but different from that obtained from spin-wave fittings~\cite{Shen2019}.

\textit{Discussion and Outlook.---}
The determination of effective spin Hamiltonian paves 
the way towards understanding the exotic quantum states 
and phenomena, as well as designing future quantum 
applications, of the quantum magnetic materials.
Solving the quantum many-body problem, 
i.e., computing the ground-state, thermodynamics, 
and dynamical properties from a spin lattice model 
constitutes a challenging problem. Therefore, 
at a first glance, the inverse problem --- learning 
the microscopic model from macroscopic 
measurements --- seems a problem intractable. 
Here, we show, through solving the artificial and 
realistic problems, that the inverse many-body problem 
can be elegantly resolved by combining the 
thermodynamics many-body solvers and Bayesian
optimization. 

The secrete lies in the fact that we actually 
do not need to solve a full many-body problem, 
but a much only a finite-temperature one that is 
numerically much easier to compute. Therefore, 
we find the ED solver that only accesses rather 
high-$T$ regime can already find the valuable 
interaction information, when combined with Bayesian 
optimization. Furthermore, with the powerful 
thermal tensor network method as a low-$T$ solver, 
a significantly improved resolution in Hamiltonian
parameters can be obtained.

Our approach, in particular when combining the 
thermal tensor network approach and Bayesian optimization, 
can provide a very promising tool in 
studying quantum magnets and uncovering 
novel quantum states and phases therein. 
For example, the family of rare-earth 
Chalcogenides AReCh$_2$ (A for alkali or 
monovalent ions, Re is rare earth, 
and Ch is O, S, or Se)~\cite{Liu2018,Zhang2020effective} 
shares a similar class of Hamiltonians
with different coupling parameters. 
As there are abundant experimental thermodynamics 
data available, the approach established here 
allows us to search for the most promising 
quantum spin liquid candidates.
Moreover, it also gives us the hope to build up a 
quantum magnetism genome library, by automatically 
finding the effective spin Hamiltonians for quantum 
magnetic materials, which are important for their future 
applications as, e.g., quantum critical coolant~\cite{Zhitomirsky2003,Zhitomirsky2004,
Garst2005,Wolf2011,Gegenwart2016} and spin-chain
quantum information data bus~\cite{Karbach2005,Cappellaro2007}, etc. 
With the automatic Hamiltonian searching framework offered,
and proof-of-principle examples tested, all these exciting 
exploration of correlated quantum materials can be started from here.

\textit{Acknowledgments.---}
W.L. thanks Shi-Ju Ran for the introduction to
active learning and Bayesian optimization, 
and Lei Wang for stimulating discussions
on the automatic differentiation.
This work was supported by the NSFC
through Grant Nos. 11974036 and 11834014. 
Source code relevant to this work and an interactive demo are 
available at \href{https://github.com/QMagen}{this https URL}.

\AtEndEnvironment{thebibliography}{
\bibitem{SM} In Supplementary Materials, we summarize three algorithms adopted in our Hamiltonian searching in 
Sec.~{\color{blue}A}. Automatic differentiation and Bayesian optimization are briefly recapitulated in Sec.~{\color{blue}B} and 
Sec.~{\color{blue}C} respectively. We also revisit the basic idea of quantum many-body methods used in this work in Sec.~{\color{blue}D}.
More fitting data on the XXZ HAFC and TMGO systems are presented in Sec.~{\color{blue}E} and Sec.~{\color{blue}F}.
}

\bibliography{QMagenRef}

\begin{thebibliography}{59}%
\makeatletter
\providecommand \@ifxundefined [1]{%
 \@ifx{#1\undefined}
}%
\providecommand \@ifnum [1]{%
 \ifnum #1\expandafter \@firstoftwo
 \else \expandafter \@secondoftwo
 \fi
}%
\providecommand \@ifx [1]{%
 \ifx #1\expandafter \@firstoftwo
 \else \expandafter \@secondoftwo
 \fi
}%
\providecommand \natexlab [1]{#1}%
\providecommand \enquote  [1]{``#1''}%
\providecommand \bibnamefont  [1]{#1}%
\providecommand \bibfnamefont [1]{#1}%
\providecommand \citenamefont [1]{#1}%
\providecommand \href@noop [0]{\@secondoftwo}%
\providecommand \href [0]{\begingroup \@sanitize@url \@href}%
\providecommand \@href[1]{\@@startlink{#1}\@@href}%
\providecommand \@@href[1]{\endgroup#1\@@endlink}%
\providecommand \@sanitize@url [0]{\catcode `\\12\catcode `\$12\catcode
  `\&12\catcode `\#12\catcode `\^12\catcode `\_12\catcode `\%12\relax}%
\providecommand \@@startlink[1]{}%
\providecommand \@@endlink[0]{}%
\providecommand \url  [0]{\begingroup\@sanitize@url \@url }%
\providecommand \@url [1]{\endgroup\@href {#1}{\urlprefix }}%
\providecommand \urlprefix  [0]{URL }%
\providecommand \Eprint [0]{\href }%
\providecommand \doibase [0]{http://dx.doi.org/}%
\providecommand \selectlanguage [0]{\@gobble}%
\providecommand \bibinfo  [0]{\@secondoftwo}%
\providecommand \bibfield  [0]{\@secondoftwo}%
\providecommand \translation [1]{[#1]}%
\providecommand \BibitemOpen [0]{}%
\providecommand \bibitemStop [0]{}%
\providecommand \bibitemNoStop [0]{.\EOS\space}%
\providecommand \EOS [0]{\spacefactor3000\relax}%
\providecommand \BibitemShut  [1]{\csname bibitem#1\endcsname}%
\let\auto@bib@innerbib\@empty
\bibitem [{\citenamefont {Anderson}(1973)}]{Anderson1973}%
  \BibitemOpen
  \bibfield  {author} {\bibinfo {author} {\bibfnamefont {P.~W.}\ \bibnamefont
  {Anderson}},\ }\bibfield  {title} {\enquote {\bibinfo {title} {Resonating
  valence bonds: A new kind of insulator?}}\ }\href {\doibase
  https://doi.org/10.1016/0025-5408(73)90167-0} {\bibfield  {journal} {\bibinfo
   {journal} {Mater. Res. Bull.}\ }\textbf {\bibinfo {volume} {8}},\ \bibinfo
  {pages} {153 -- 160} (\bibinfo {year} {1973})}\BibitemShut {NoStop}%
\bibitem [{\citenamefont {Kitaev}(2006)}]{Kitaev2006}%
  \BibitemOpen
  \bibfield  {author} {\bibinfo {author} {\bibfnamefont {A.}~\bibnamefont
  {Kitaev}},\ }\bibfield  {title} {\enquote {\bibinfo {title} {Anyons in an
  exactly solved model and beyond},}\ }\href {\doibase
  https://doi.org/10.1016/j.aop.2005.10.005} {\bibfield  {journal} {\bibinfo
  {journal} {Ann. Phys.}\ }\textbf {\bibinfo {volume} {321}},\ \bibinfo {pages}
  {2 -- 111} (\bibinfo {year} {2006})},\ \bibinfo {note} {{January Special
  Issue}}\BibitemShut {NoStop}%
\bibitem [{\citenamefont {Zhou}\ \emph {et~al.}(2017)\citenamefont {Zhou},
  \citenamefont {Kanoda},\ and\ \citenamefont {Ng}}]{Zhou2017}%
  \BibitemOpen
  \bibfield  {author} {\bibinfo {author} {\bibfnamefont {Y.}~\bibnamefont
  {Zhou}}, \bibinfo {author} {\bibfnamefont {K.}~\bibnamefont {Kanoda}}, \ and\
  \bibinfo {author} {\bibfnamefont {T.-K.}\ \bibnamefont {Ng}},\ }\bibfield
  {title} {\enquote {\bibinfo {title} {Quantum spin liquid states},}\ }\href
  {\doibase 10.1103/RevModPhys.89.025003} {\bibfield  {journal} {\bibinfo
  {journal} {Rev. Mod. Phys.}\ }\textbf {\bibinfo {volume} {89}},\ \bibinfo
  {pages} {025003} (\bibinfo {year} {2017})}\BibitemShut {NoStop}%
\bibitem [{\citenamefont {{Balents}}(2010)}]{Balents2010}%
  \BibitemOpen
  \bibfield  {author} {\bibinfo {author} {\bibfnamefont {L.}~\bibnamefont
  {{Balents}}},\ }\bibfield  {title} {\enquote {\bibinfo {title} {{Spin liquids
  in frustrated magnets}},}\ }\href {\doibase 10.1038/nature08917} {\bibfield
  {journal} {\bibinfo  {journal} {\nat}\ }\textbf {\bibinfo {volume} {464}},\
  \bibinfo {pages} {199--208} (\bibinfo {year} {2010})}\BibitemShut {NoStop}%
\bibitem [{\citenamefont {Han}\ \emph {et~al.}(2012)\citenamefont {Han},
  \citenamefont {Helton}, \citenamefont {Chu}, \citenamefont {Nocera},
  \citenamefont {Rodriguez-Rivera}, \citenamefont {Broholm},\ and\
  \citenamefont {Lee}}]{Han2012}%
  \BibitemOpen
  \bibfield  {author} {\bibinfo {author} {\bibfnamefont {T.~H.}\ \bibnamefont
  {Han}}, \bibinfo {author} {\bibfnamefont {J.~S.}\ \bibnamefont {Helton}},
  \bibinfo {author} {\bibfnamefont {S.}~\bibnamefont {Chu}}, \bibinfo {author}
  {\bibfnamefont {D.~G.}\ \bibnamefont {Nocera}}, \bibinfo {author}
  {\bibfnamefont {J.~A.}\ \bibnamefont {Rodriguez-Rivera}}, \bibinfo {author}
  {\bibfnamefont {C.}~\bibnamefont {Broholm}}, \ and\ \bibinfo {author}
  {\bibfnamefont {Y.~S.}\ \bibnamefont {Lee}},\ }\bibfield  {title} {\enquote
  {\bibinfo {title} {Fractionalized excitations in the spin-liquid state of a
  kagome-lattice antiferromagnet},}\ }\href
  {https://doi.org/10.1038/nature11659} {\bibfield  {journal} {\bibinfo
  {journal} {Nature}\ }\textbf {\bibinfo {volume} {492}},\ \bibinfo {pages}
  {406--410} (\bibinfo {year} {2012})}\BibitemShut {NoStop}%
\bibitem [{\citenamefont {Fu}\ \emph {et~al.}(2015)\citenamefont {Fu},
  \citenamefont {Imai}, \citenamefont {Han},\ and\ \citenamefont
  {Lee}}]{Fu2015}%
  \BibitemOpen
  \bibfield  {author} {\bibinfo {author} {\bibfnamefont {Mingxuan}\
  \bibnamefont {Fu}}, \bibinfo {author} {\bibfnamefont {Takashi}\ \bibnamefont
  {Imai}}, \bibinfo {author} {\bibfnamefont {Tian-Heng}\ \bibnamefont {Han}}, \
  and\ \bibinfo {author} {\bibfnamefont {Young~S.}\ \bibnamefont {Lee}},\
  }\bibfield  {title} {\enquote {\bibinfo {title} {Evidence for a gapped
  spin-liquid ground state in a kagome {Heisenberg} antiferromagnet},}\ }\href
  {https://science.sciencemag.org/content/350/6261/655} {\bibfield  {journal}
  {\bibinfo  {journal} {Science}\ }\textbf {\bibinfo {volume} {350}},\ \bibinfo
  {pages} {655--658} (\bibinfo {year} {2015})}\BibitemShut {NoStop}%
\bibitem [{\citenamefont {Shimizu}\ \emph {et~al.}(2003)\citenamefont
  {Shimizu}, \citenamefont {Miyagawa}, \citenamefont {Kanoda}, \citenamefont
  {Maesato},\ and\ \citenamefont {Saito}}]{Shimizu2003}%
  \BibitemOpen
  \bibfield  {author} {\bibinfo {author} {\bibfnamefont {Y.}~\bibnamefont
  {Shimizu}}, \bibinfo {author} {\bibfnamefont {K.}~\bibnamefont {Miyagawa}},
  \bibinfo {author} {\bibfnamefont {K.}~\bibnamefont {Kanoda}}, \bibinfo
  {author} {\bibfnamefont {M.}~\bibnamefont {Maesato}}, \ and\ \bibinfo
  {author} {\bibfnamefont {G.}~\bibnamefont {Saito}},\ }\bibfield  {title}
  {\enquote {\bibinfo {title} {Spin liquid state in an organic {Mott} insulator
  with a triangular lattice},}\ }\href {\doibase 10.1103/PhysRevLett.91.107001}
  {\bibfield  {journal} {\bibinfo  {journal} {Phys. Rev. Lett.}\ }\textbf
  {\bibinfo {volume} {91}},\ \bibinfo {pages} {107001} (\bibinfo {year}
  {2003})}\BibitemShut {NoStop}%
\bibitem [{\citenamefont {{Yamashita}}\ \emph {et~al.}(2010)\citenamefont
  {{Yamashita}}, \citenamefont {{Nakata}}, \citenamefont {{Senshu}},
  \citenamefont {{Nagata}}, \citenamefont {{Yamamoto}}, \citenamefont {{Kato}},
  \citenamefont {{Shibauchi}},\ and\ \citenamefont
  {{Matsuda}}}]{Yamashita2010}%
  \BibitemOpen
  \bibfield  {author} {\bibinfo {author} {\bibfnamefont {M.}~\bibnamefont
  {{Yamashita}}}, \bibinfo {author} {\bibfnamefont {N.}~\bibnamefont
  {{Nakata}}}, \bibinfo {author} {\bibfnamefont {Y.}~\bibnamefont {{Senshu}}},
  \bibinfo {author} {\bibfnamefont {M.}~\bibnamefont {{Nagata}}}, \bibinfo
  {author} {\bibfnamefont {H.~M.}\ \bibnamefont {{Yamamoto}}}, \bibinfo
  {author} {\bibfnamefont {R.}~\bibnamefont {{Kato}}}, \bibinfo {author}
  {\bibfnamefont {T.}~\bibnamefont {{Shibauchi}}}, \ and\ \bibinfo {author}
  {\bibfnamefont {Y.}~\bibnamefont {{Matsuda}}},\ }\bibfield  {title} {\enquote
  {\bibinfo {title} {Highly mobile gapless excitations in a two-dimensional
  candidate quantum spin liquid},}\ }\href {\doibase 10.1126/science.1188200}
  {\bibfield  {journal} {\bibinfo  {journal} {Science}\ }\textbf {\bibinfo
  {volume} {328}},\ \bibinfo {pages} {1246} (\bibinfo {year}
  {2010})}\BibitemShut {NoStop}%
\bibitem [{\citenamefont {Liu}\ \emph {et~al.}(2018)\citenamefont {Liu},
  \citenamefont {Zhang}, \citenamefont {Ji}, \citenamefont {Liu}, \citenamefont
  {Li}, \citenamefont {Wang}, \citenamefont {Lei}, \citenamefont {Chen},\ and\
  \citenamefont {Zhang}}]{Liu2018}%
  \BibitemOpen
  \bibfield  {author} {\bibinfo {author} {\bibfnamefont {Weiwei}\ \bibnamefont
  {Liu}}, \bibinfo {author} {\bibfnamefont {Zheng}\ \bibnamefont {Zhang}},
  \bibinfo {author} {\bibfnamefont {Jianting}\ \bibnamefont {Ji}}, \bibinfo
  {author} {\bibfnamefont {Yixuan}\ \bibnamefont {Liu}}, \bibinfo {author}
  {\bibfnamefont {Jianshu}\ \bibnamefont {Li}}, \bibinfo {author}
  {\bibfnamefont {Xiaoqun}\ \bibnamefont {Wang}}, \bibinfo {author}
  {\bibfnamefont {Hechang}\ \bibnamefont {Lei}}, \bibinfo {author}
  {\bibfnamefont {Gang}\ \bibnamefont {Chen}}, \ and\ \bibinfo {author}
  {\bibfnamefont {Qingming}\ \bibnamefont {Zhang}},\ }\bibfield  {title}
  {\enquote {\bibinfo {title} {Rare-earth chalcogenides: A large family of
  triangular lattice spin liquid candidates},}\ }\href {\doibase
  10.1088/0256-307x/35/11/117501} {\bibfield  {journal} {\bibinfo  {journal}
  {Chinese Physics Letters}\ }\textbf {\bibinfo {volume} {35}},\ \bibinfo
  {pages} {117501} (\bibinfo {year} {2018})}\BibitemShut {NoStop}%
\bibitem [{\citenamefont {Jackeli}\ and\ \citenamefont
  {Khaliullin}(2009)}]{Jackeli2009}%
  \BibitemOpen
  \bibfield  {author} {\bibinfo {author} {\bibfnamefont {G.}~\bibnamefont
  {Jackeli}}\ and\ \bibinfo {author} {\bibfnamefont {G.}~\bibnamefont
  {Khaliullin}},\ }\bibfield  {title} {\enquote {\bibinfo {title} {Mott
  insulators in the strong spin-orbit coupling limit: From {Heisenberg} to a
  quantum compass and {Kitaev} models},}\ }\href {\doibase
  10.1103/PhysRevLett.102.017205} {\bibfield  {journal} {\bibinfo  {journal}
  {Phys. Rev. Lett.}\ }\textbf {\bibinfo {volume} {102}},\ \bibinfo {pages}
  {017205} (\bibinfo {year} {2009})}\BibitemShut {NoStop}%
\bibitem [{\citenamefont {Chaloupka}\ \emph {et~al.}(2010)\citenamefont
  {Chaloupka}, \citenamefont {Jackeli},\ and\ \citenamefont
  {Khaliullin}}]{Chaloupka2010}%
  \BibitemOpen
  \bibfield  {author} {\bibinfo {author} {\bibfnamefont {J.}~\bibnamefont
  {Chaloupka}}, \bibinfo {author} {\bibfnamefont {G.}~\bibnamefont {Jackeli}},
  \ and\ \bibinfo {author} {\bibfnamefont {G.}~\bibnamefont {Khaliullin}},\
  }\bibfield  {title} {\enquote {\bibinfo {title} {{Kitaev-Heisenberg} model on
  a honeycomb lattice: Possible exotic phases in iridium oxides
  {${A}_{2}{\mathrm{IrO}}_{3}$}},}\ }\href {\doibase
  10.1103/PhysRevLett.105.027204} {\bibfield  {journal} {\bibinfo  {journal}
  {Phys. Rev. Lett.}\ }\textbf {\bibinfo {volume} {105}},\ \bibinfo {pages}
  {027204} (\bibinfo {year} {2010})}\BibitemShut {NoStop}%
\bibitem [{\citenamefont {Ye}\ \emph {et~al.}(2012)\citenamefont {Ye},
  \citenamefont {Chi}, \citenamefont {Cao}, \citenamefont {Chakoumakos},
  \citenamefont {Fernandez-Baca}, \citenamefont {Custelcean}, \citenamefont
  {Qi}, \citenamefont {Korneta},\ and\ \citenamefont {Cao}}]{Ye2012}%
  \BibitemOpen
  \bibfield  {author} {\bibinfo {author} {\bibfnamefont {F.}~\bibnamefont
  {Ye}}, \bibinfo {author} {\bibfnamefont {S.}~\bibnamefont {Chi}}, \bibinfo
  {author} {\bibfnamefont {H.}~\bibnamefont {Cao}}, \bibinfo {author}
  {\bibfnamefont {B.~C.}\ \bibnamefont {Chakoumakos}}, \bibinfo {author}
  {\bibfnamefont {J.~A.}\ \bibnamefont {Fernandez-Baca}}, \bibinfo {author}
  {\bibfnamefont {R.}~\bibnamefont {Custelcean}}, \bibinfo {author}
  {\bibfnamefont {T.~F.}\ \bibnamefont {Qi}}, \bibinfo {author} {\bibfnamefont
  {O.~B.}\ \bibnamefont {Korneta}}, \ and\ \bibinfo {author} {\bibfnamefont
  {G.}~\bibnamefont {Cao}},\ }\bibfield  {title} {\enquote {\bibinfo {title}
  {Direct evidence of a zigzag spin-chain structure in the honeycomb lattice: A
  neutron and {X-ray} diffraction investigation of single-crystal
  {Na${}_{2}$IrO${}_{3}$}},}\ }\href {\doibase 10.1103/PhysRevB.85.180403}
  {\bibfield  {journal} {\bibinfo  {journal} {Phys. Rev. B}\ }\textbf {\bibinfo
  {volume} {85}},\ \bibinfo {pages} {180403} (\bibinfo {year}
  {2012})}\BibitemShut {NoStop}%
\bibitem [{\citenamefont {{Banerjee}}\ \emph {et~al.}(2017)\citenamefont
  {{Banerjee}}, \citenamefont {{Yan}}, \citenamefont {{Knolle}}, \citenamefont
  {{Bridges}}, \citenamefont {{Stone}}, \citenamefont {{Lumsden}},
  \citenamefont {{Mandrus}}, \citenamefont {{Tennant}}, \citenamefont
  {{Moessner}},\ and\ \citenamefont {{Nagler}}}]{Banerjee2017}%
  \BibitemOpen
  \bibfield  {author} {\bibinfo {author} {\bibfnamefont {A.}~\bibnamefont
  {{Banerjee}}}, \bibinfo {author} {\bibfnamefont {J.}~\bibnamefont {{Yan}}},
  \bibinfo {author} {\bibfnamefont {J.}~\bibnamefont {{Knolle}}}, \bibinfo
  {author} {\bibfnamefont {C.~A.}\ \bibnamefont {{Bridges}}}, \bibinfo {author}
  {\bibfnamefont {M.~B.}\ \bibnamefont {{Stone}}}, \bibinfo {author}
  {\bibfnamefont {M.~D.}\ \bibnamefont {{Lumsden}}}, \bibinfo {author}
  {\bibfnamefont {D.~G.}\ \bibnamefont {{Mandrus}}}, \bibinfo {author}
  {\bibfnamefont {D.~A.}\ \bibnamefont {{Tennant}}}, \bibinfo {author}
  {\bibfnamefont {R.}~\bibnamefont {{Moessner}}}, \ and\ \bibinfo {author}
  {\bibfnamefont {S.~E.}\ \bibnamefont {{Nagler}}},\ }\bibfield  {title}
  {\enquote {\bibinfo {title} {{Neutron scattering in the proximate quantum
  spin liquid {\ensuremath{\alpha}}-RuCl$_{3}$}},}\ }\href {\doibase
  10.1126/science.aah6015} {\bibfield  {journal} {\bibinfo  {journal}
  {Science}\ }\textbf {\bibinfo {volume} {356}},\ \bibinfo {pages} {1055--1059}
  (\bibinfo {year} {2017})}\BibitemShut {NoStop}%
\bibitem [{\citenamefont {Laurell}\ and\ \citenamefont
  {Okamoto}(2020)}]{Laurell2020}%
  \BibitemOpen
  \bibfield  {author} {\bibinfo {author} {\bibfnamefont {Pontus}\ \bibnamefont
  {Laurell}}\ and\ \bibinfo {author} {\bibfnamefont {Satoshi}\ \bibnamefont
  {Okamoto}},\ }\bibfield  {title} {\enquote {\bibinfo {title} {Dynamical and
  thermal magnetic properties of the {Kitaev} spin liquid candidate
  $\alpha$-{RuCl3}},}\ }\href@noop {} {\bibfield  {journal} {\bibinfo
  {journal} {npj Quantum Materials}\ }\textbf {\bibinfo {volume} {5}},\
  \bibinfo {pages} {2} (\bibinfo {year} {2020})}\BibitemShut {NoStop}%
\bibitem [{\citenamefont {Bursill}\ \emph {et~al.}(1996)\citenamefont
  {Bursill}, \citenamefont {Xiang},\ and\ \citenamefont
  {Gehring}}]{Bursill1996}%
  \BibitemOpen
  \bibfield  {author} {\bibinfo {author} {\bibfnamefont {R.~J.}\ \bibnamefont
  {Bursill}}, \bibinfo {author} {\bibfnamefont {T.}~\bibnamefont {Xiang}}, \
  and\ \bibinfo {author} {\bibfnamefont {G.~A.}\ \bibnamefont {Gehring}},\
  }\bibfield  {title} {\enquote {\bibinfo {title} {The density matrix
  renormalization group for a quantum spin chain at non-zero temperature},}\
  }\href {http://stacks.iop.org/0953-8984/8/i=40/a=003} {\bibfield  {journal}
  {\bibinfo  {journal} {J. Phys. Condens.}\ }\textbf {\bibinfo {volume} {8}},\
  \bibinfo {pages} {L583} (\bibinfo {year} {1996})}\BibitemShut {NoStop}%
\bibitem [{\citenamefont {Wang}\ and\ \citenamefont {Xiang}(1997)}]{Wang1997}%
  \BibitemOpen
  \bibfield  {author} {\bibinfo {author} {\bibfnamefont {X.}~\bibnamefont
  {Wang}}\ and\ \bibinfo {author} {\bibfnamefont {T.}~\bibnamefont {Xiang}},\
  }\bibfield  {title} {\enquote {\bibinfo {title} {Transfer-matrix
  density-matrix renormalization-group theory for thermodynamics of
  one-dimensional quantum systems},}\ }\href {\doibase
  10.1103/PhysRevB.56.5061} {\bibfield  {journal} {\bibinfo  {journal} {Phys.
  Rev. B}\ }\textbf {\bibinfo {volume} {56}},\ \bibinfo {pages} {5061--5064}
  (\bibinfo {year} {1997})}\BibitemShut {NoStop}%
\bibitem [{\citenamefont {Xiang}(1998)}]{Xiang1998}%
  \BibitemOpen
  \bibfield  {author} {\bibinfo {author} {\bibfnamefont {T.}~\bibnamefont
  {Xiang}},\ }\bibfield  {title} {\enquote {\bibinfo {title} {Thermodynamics of
  quantum {Heisenberg} spin chains},}\ }\href {\doibase
  10.1103/PhysRevB.58.9142} {\bibfield  {journal} {\bibinfo  {journal} {Phys.
  Rev. B}\ }\textbf {\bibinfo {volume} {58}},\ \bibinfo {pages} {9142--9149}
  (\bibinfo {year} {1998})}\BibitemShut {NoStop}%
\bibitem [{\citenamefont {Feiguin}\ and\ \citenamefont
  {White}(2005)}]{Feiguin2005}%
  \BibitemOpen
  \bibfield  {author} {\bibinfo {author} {\bibfnamefont {A.~E.}\ \bibnamefont
  {Feiguin}}\ and\ \bibinfo {author} {\bibfnamefont {S.~R.}\ \bibnamefont
  {White}},\ }\bibfield  {title} {\enquote {\bibinfo {title}
  {Finite-temperature density matrix renormalization using an enlarged
  {Hilbert} space},}\ }\href {\doibase 10.1103/PhysRevB.72.220401} {\bibfield
  {journal} {\bibinfo  {journal} {Phys. Rev. B}\ }\textbf {\bibinfo {volume}
  {72}},\ \bibinfo {pages} {220401(R)} (\bibinfo {year} {2005})}\BibitemShut
  {NoStop}%
\bibitem [{\citenamefont {White}(2009)}]{White2009}%
  \BibitemOpen
  \bibfield  {author} {\bibinfo {author} {\bibfnamefont {S.~R.}\ \bibnamefont
  {White}},\ }\bibfield  {title} {\enquote {\bibinfo {title} {Minimally
  entangled typical quantum states at finite temperature},}\ }\href {\doibase
  10.1103/PhysRevLett.102.190601} {\bibfield  {journal} {\bibinfo  {journal}
  {Phys. Rev. Lett.}\ }\textbf {\bibinfo {volume} {102}},\ \bibinfo {pages}
  {190601} (\bibinfo {year} {2009})}\BibitemShut {NoStop}%
\bibitem [{\citenamefont {Stoudenmire}\ and\ \citenamefont
  {White}(2010)}]{Stoudemire2010}%
  \BibitemOpen
  \bibfield  {author} {\bibinfo {author} {\bibfnamefont {E.~M.}\ \bibnamefont
  {Stoudenmire}}\ and\ \bibinfo {author} {\bibfnamefont {S.~R.}\ \bibnamefont
  {White}},\ }\bibfield  {title} {\enquote {\bibinfo {title} {Minimally
  entangled typical thermal state algorithms},}\ }\href
  {http://stacks.iop.org/1367-2630/12/i=5/a=055026} {\bibfield  {journal}
  {\bibinfo  {journal} {New J. Phys.}\ }\textbf {\bibinfo {volume} {12}},\
  \bibinfo {pages} {055026} (\bibinfo {year} {2010})}\BibitemShut {NoStop}%
\bibitem [{\citenamefont {Li}\ \emph {et~al.}(2011)\citenamefont {Li},
  \citenamefont {Ran}, \citenamefont {Gong}, \citenamefont {Zhao},
  \citenamefont {Xi}, \citenamefont {Ye},\ and\ \citenamefont {Su}}]{Li2011}%
  \BibitemOpen
  \bibfield  {author} {\bibinfo {author} {\bibfnamefont {W.}~\bibnamefont
  {Li}}, \bibinfo {author} {\bibfnamefont {S.-J.}\ \bibnamefont {Ran}},
  \bibinfo {author} {\bibfnamefont {S.-S.}\ \bibnamefont {Gong}}, \bibinfo
  {author} {\bibfnamefont {Y.}~\bibnamefont {Zhao}}, \bibinfo {author}
  {\bibfnamefont {B.}~\bibnamefont {Xi}}, \bibinfo {author} {\bibfnamefont
  {F.}~\bibnamefont {Ye}}, \ and\ \bibinfo {author} {\bibfnamefont
  {G.}~\bibnamefont {Su}},\ }\bibfield  {title} {\enquote {\bibinfo {title}
  {Linearized tensor renormalization group algorithm for the calculation of
  thermodynamic properties of quantum lattice models},}\ }\href {\doibase
  10.1103/PhysRevLett.106.127202} {\bibfield  {journal} {\bibinfo  {journal}
  {Phys. Rev. Lett.}\ }\textbf {\bibinfo {volume} {106}},\ \bibinfo {pages}
  {127202} (\bibinfo {year} {2011})}\BibitemShut {NoStop}%
\bibitem [{\citenamefont {Dong}\ \emph {et~al.}(2017)\citenamefont {Dong},
  \citenamefont {Chen}, \citenamefont {Liu},\ and\ \citenamefont
  {Li}}]{Dong2017}%
  \BibitemOpen
  \bibfield  {author} {\bibinfo {author} {\bibfnamefont {Y.-L.}\ \bibnamefont
  {Dong}}, \bibinfo {author} {\bibfnamefont {L.}~\bibnamefont {Chen}}, \bibinfo
  {author} {\bibfnamefont {Y.-J.}\ \bibnamefont {Liu}}, \ and\ \bibinfo
  {author} {\bibfnamefont {W.}~\bibnamefont {Li}},\ }\bibfield  {title}
  {\enquote {\bibinfo {title} {Bilayer linearized tensor renormalization group
  approach for thermal tensor networks},}\ }\href {\doibase
  10.1103/PhysRevB.95.144428} {\bibfield  {journal} {\bibinfo  {journal} {Phys.
  Rev. B}\ }\textbf {\bibinfo {volume} {95}},\ \bibinfo {pages} {144428}
  (\bibinfo {year} {2017})}\BibitemShut {NoStop}%
\bibitem [{\citenamefont {Chen}\ \emph {et~al.}(2017)\citenamefont {Chen},
  \citenamefont {Liu}, \citenamefont {Chen},\ and\ \citenamefont
  {Li}}]{Chen2017}%
  \BibitemOpen
  \bibfield  {author} {\bibinfo {author} {\bibfnamefont {B.-B.}\ \bibnamefont
  {Chen}}, \bibinfo {author} {\bibfnamefont {Y.-J.}\ \bibnamefont {Liu}},
  \bibinfo {author} {\bibfnamefont {Z.}~\bibnamefont {Chen}}, \ and\ \bibinfo
  {author} {\bibfnamefont {W.}~\bibnamefont {Li}},\ }\bibfield  {title}
  {\enquote {\bibinfo {title} {Series-expansion thermal tensor network approach
  for quantum lattice models},}\ }\href {\doibase 10.1103/PhysRevB.95.161104}
  {\bibfield  {journal} {\bibinfo  {journal} {Phys. Rev. B}\ }\textbf {\bibinfo
  {volume} {95}},\ \bibinfo {pages} {161104(R)} (\bibinfo {year}
  {2017})}\BibitemShut {NoStop}%
\bibitem [{\citenamefont {Chen}\ \emph {et~al.}(2018)\citenamefont {Chen},
  \citenamefont {Chen}, \citenamefont {Chen}, \citenamefont {Li},\ and\
  \citenamefont {Weichselbaum}}]{Chen2018}%
  \BibitemOpen
  \bibfield  {author} {\bibinfo {author} {\bibfnamefont {B.-B.}\ \bibnamefont
  {Chen}}, \bibinfo {author} {\bibfnamefont {L.}~\bibnamefont {Chen}}, \bibinfo
  {author} {\bibfnamefont {Z.}~\bibnamefont {Chen}}, \bibinfo {author}
  {\bibfnamefont {W.}~\bibnamefont {Li}}, \ and\ \bibinfo {author}
  {\bibfnamefont {A.}~\bibnamefont {Weichselbaum}},\ }\bibfield  {title}
  {\enquote {\bibinfo {title} {Exponential thermal tensor network approach for
  quantum lattice models},}\ }\href {\doibase 10.1103/PhysRevX.8.031082}
  {\bibfield  {journal} {\bibinfo  {journal} {Phys. Rev. X}\ }\textbf {\bibinfo
  {volume} {8}},\ \bibinfo {pages} {031082} (\bibinfo {year}
  {2018})}\BibitemShut {NoStop}%
\bibitem [{\citenamefont {Li}\ \emph {et~al.}(2019)\citenamefont {Li},
  \citenamefont {Chen}, \citenamefont {Chen}, \citenamefont {von Delft},
  \citenamefont {Weichselbaum},\ and\ \citenamefont {Li}}]{Lih2019}%
  \BibitemOpen
  \bibfield  {author} {\bibinfo {author} {\bibfnamefont {H.}~\bibnamefont
  {Li}}, \bibinfo {author} {\bibfnamefont {B.-B.}\ \bibnamefont {Chen}},
  \bibinfo {author} {\bibfnamefont {Z.}~\bibnamefont {Chen}}, \bibinfo {author}
  {\bibfnamefont {J.}~\bibnamefont {von Delft}}, \bibinfo {author}
  {\bibfnamefont {A.}~\bibnamefont {Weichselbaum}}, \ and\ \bibinfo {author}
  {\bibfnamefont {W.}~\bibnamefont {Li}},\ }\bibfield  {title} {\enquote
  {\bibinfo {title} {Thermal tensor renormalization group simulations of
  square-lattice quantum spin models},}\ }\href {\doibase
  10.1103/PhysRevB.100.045110} {\bibfield  {journal} {\bibinfo  {journal}
  {Phys. Rev. B}\ }\textbf {\bibinfo {volume} {100}},\ \bibinfo {pages}
  {045110} (\bibinfo {year} {2019})}\BibitemShut {NoStop}%
\bibitem [{\citenamefont {Carleo}\ and\ \citenamefont
  {Troyer}(2017)}]{Carleo2017}%
  \BibitemOpen
  \bibfield  {author} {\bibinfo {author} {\bibfnamefont {G.}~\bibnamefont
  {Carleo}}\ and\ \bibinfo {author} {\bibfnamefont {M.}~\bibnamefont
  {Troyer}},\ }\bibfield  {title} {\enquote {\bibinfo {title} {Solving the
  quantum many-body problem with artificial neural networks},}\ }\href
  {\doibase 10.1126/science.aag2302} {\bibfield  {journal} {\bibinfo  {journal}
  {Science}\ }\textbf {\bibinfo {volume} {355}},\ \bibinfo {pages} {602--606}
  (\bibinfo {year} {2017})}\BibitemShut {NoStop}%
\bibitem [{\citenamefont {Liao}\ \emph {et~al.}(2019)\citenamefont {Liao},
  \citenamefont {Liu}, \citenamefont {Wang},\ and\ \citenamefont
  {Xiang}}]{Liao2019}%
  \BibitemOpen
  \bibfield  {author} {\bibinfo {author} {\bibfnamefont {H.-J.}\ \bibnamefont
  {Liao}}, \bibinfo {author} {\bibfnamefont {J.-G.}\ \bibnamefont {Liu}},
  \bibinfo {author} {\bibfnamefont {L.}~\bibnamefont {Wang}}, \ and\ \bibinfo
  {author} {\bibfnamefont {T.}~\bibnamefont {Xiang}},\ }\bibfield  {title}
  {\enquote {\bibinfo {title} {Differentiable programming tensor networks},}\
  }\href {\doibase 10.1103/PhysRevX.9.031041} {\bibfield  {journal} {\bibinfo
  {journal} {Phys. Rev. X}\ }\textbf {\bibinfo {volume} {9}},\ \bibinfo {pages}
  {031041} (\bibinfo {year} {2019})}\BibitemShut {NoStop}%
\bibitem [{\citenamefont {Chen}\ \emph {et~al.}(2020)\citenamefont {Chen},
  \citenamefont {Gao}, \citenamefont {Guo}, \citenamefont {Liu}, \citenamefont
  {Zhao}, \citenamefont {Liao}, \citenamefont {Wang}, \citenamefont {Xiang},
  \citenamefont {Li},\ and\ \citenamefont {Xie}}]{Chen2020}%
  \BibitemOpen
  \bibfield  {author} {\bibinfo {author} {\bibfnamefont {Bin-Bin}\ \bibnamefont
  {Chen}}, \bibinfo {author} {\bibfnamefont {Yuan}\ \bibnamefont {Gao}},
  \bibinfo {author} {\bibfnamefont {Yi-Bin}\ \bibnamefont {Guo}}, \bibinfo
  {author} {\bibfnamefont {Yuzhi}\ \bibnamefont {Liu}}, \bibinfo {author}
  {\bibfnamefont {Hui-Hai}\ \bibnamefont {Zhao}}, \bibinfo {author}
  {\bibfnamefont {Hai-Jun}\ \bibnamefont {Liao}}, \bibinfo {author}
  {\bibfnamefont {Lei}\ \bibnamefont {Wang}}, \bibinfo {author} {\bibfnamefont
  {Tao}\ \bibnamefont {Xiang}}, \bibinfo {author} {\bibfnamefont {Wei}\
  \bibnamefont {Li}}, \ and\ \bibinfo {author} {\bibfnamefont {Z.~Y.}\
  \bibnamefont {Xie}},\ }\bibfield  {title} {\enquote {\bibinfo {title}
  {Automatic differentiation for second renormalization of tensor networks},}\
  }\href {\doibase 10.1103/PhysRevB.101.220409} {\bibfield  {journal} {\bibinfo
   {journal} {Phys. Rev. B}\ }\textbf {\bibinfo {volume} {101}},\ \bibinfo
  {pages} {220409} (\bibinfo {year} {2020})}\BibitemShut {NoStop}%
\bibitem [{\citenamefont {Stoudenmire}\ and\ \citenamefont
  {Schwab}(2016)}]{Stoudenmire2016}%
  \BibitemOpen
  \bibfield  {author} {\bibinfo {author} {\bibfnamefont {E.}~\bibnamefont
  {Stoudenmire}}\ and\ \bibinfo {author} {\bibfnamefont {D.~J.}\ \bibnamefont
  {Schwab}},\ }\bibfield  {title} {\enquote {\bibinfo {title} {Supervised
  learning with tensor networks},}\ }in\ \href
  {http://papers.nips.cc/paper/6211-supervised-learning-with-tensor-networks.pdf}
  {\emph {\bibinfo {booktitle} {Advances in Neural Information Processing
  Systems 29}}},\ \bibinfo {editor} {edited by\ \bibinfo {editor}
  {\bibfnamefont {D.~D.}\ \bibnamefont {Lee}}, \bibinfo {editor} {\bibfnamefont
  {M.}~\bibnamefont {Sugiyama}}, \bibinfo {editor} {\bibfnamefont {U.~V.}\
  \bibnamefont {Luxburg}}, \bibinfo {editor} {\bibfnamefont {I.}~\bibnamefont
  {Guyon}}, \ and\ \bibinfo {editor} {\bibfnamefont {R.}~\bibnamefont
  {Garnett}}}\ (\bibinfo  {publisher} {Curran Associates, Inc.},\ \bibinfo
  {year} {2016})\ pp.\ \bibinfo {pages} {4799--4807}\BibitemShut {NoStop}%
\bibitem [{\citenamefont {Liu}\ \emph {et~al.}(2019)\citenamefont {Liu},
  \citenamefont {Ran}, \citenamefont {Wittek}, \citenamefont {Peng},
  \citenamefont {Garc{\'{\i}}a}, \citenamefont {Su},\ and\ \citenamefont
  {Lewenstein}}]{Liu2019}%
  \BibitemOpen
  \bibfield  {author} {\bibinfo {author} {\bibfnamefont {Ding}\ \bibnamefont
  {Liu}}, \bibinfo {author} {\bibfnamefont {Shi-Ju}\ \bibnamefont {Ran}},
  \bibinfo {author} {\bibfnamefont {Peter}\ \bibnamefont {Wittek}}, \bibinfo
  {author} {\bibfnamefont {Cheng}\ \bibnamefont {Peng}}, \bibinfo {author}
  {\bibfnamefont {Raul~Bl{\'{a}}zquez}\ \bibnamefont {Garc{\'{\i}}a}}, \bibinfo
  {author} {\bibfnamefont {Gang}\ \bibnamefont {Su}}, \ and\ \bibinfo {author}
  {\bibfnamefont {Maciej}\ \bibnamefont {Lewenstein}},\ }\bibfield  {title}
  {\enquote {\bibinfo {title} {Machine learning by unitary tensor network of
  hierarchical tree structure},}\ }\href {\doibase 10.1088/1367-2630/ab31ef}
  {\bibfield  {journal} {\bibinfo  {journal} {New Journal of Physics}\ }\textbf
  {\bibinfo {volume} {21}},\ \bibinfo {pages} {073059} (\bibinfo {year}
  {2019})}\BibitemShut {NoStop}%
\bibitem [{\citenamefont {Ran}(2020)}]{Ran2020}%
  \BibitemOpen
  \bibfield  {author} {\bibinfo {author} {\bibfnamefont {Shi-Ju}\ \bibnamefont
  {Ran}},\ }\href@noop {} {\enquote {\bibinfo {title} {Bayesian tensor network
  with polynomial complexity for probabilistic machine learning},}\ } (\bibinfo
  {year} {2020}),\ \Eprint {http://arxiv.org/abs/1912.12923} {arXiv:1912.12923
  [stat.ML]} \BibitemShut {NoStop}%
\bibitem [{\citenamefont {Cichocki}\ \emph {et~al.}(2017)\citenamefont
  {Cichocki}, \citenamefont {Phan}, \citenamefont {Zhao}, \citenamefont {Lee},
  \citenamefont {Oseledets}, \citenamefont {Sugiyama},\ and\ \citenamefont
  {Mandic}}]{Cichocki2017}%
  \BibitemOpen
  \bibfield  {author} {\bibinfo {author} {\bibfnamefont {Andrzej}\ \bibnamefont
  {Cichocki}}, \bibinfo {author} {\bibfnamefont {Anh-Huy}\ \bibnamefont
  {Phan}}, \bibinfo {author} {\bibfnamefont {Qibin}\ \bibnamefont {Zhao}},
  \bibinfo {author} {\bibfnamefont {Namgil}\ \bibnamefont {Lee}}, \bibinfo
  {author} {\bibfnamefont {Ivan}\ \bibnamefont {Oseledets}}, \bibinfo {author}
  {\bibfnamefont {Masashi}\ \bibnamefont {Sugiyama}}, \ and\ \bibinfo {author}
  {\bibfnamefont {Danilo~P.}\ \bibnamefont {Mandic}},\ }\bibfield  {title}
  {\enquote {\bibinfo {title} {Tensor networks for dimensionality reduction and
  large-scale optimization: Part 2 applications and future perspectives},}\
  }\href {\doibase 10.1561/2200000067} {\bibfield  {journal} {\bibinfo
  {journal} {Foundations and Trends® in Machine Learning}\ }\textbf {\bibinfo
  {volume} {9}},\ \bibinfo {pages} {431--673} (\bibinfo {year}
  {2017})}\BibitemShut {NoStop}%
\bibitem [{\citenamefont {Han}\ \emph {et~al.}(2018)\citenamefont {Han},
  \citenamefont {Wang}, \citenamefont {Fan}, \citenamefont {Wang},\ and\
  \citenamefont {Zhang}}]{Han2018}%
  \BibitemOpen
  \bibfield  {author} {\bibinfo {author} {\bibfnamefont {Zhao-Yu}\ \bibnamefont
  {Han}}, \bibinfo {author} {\bibfnamefont {Jun}\ \bibnamefont {Wang}},
  \bibinfo {author} {\bibfnamefont {Heng}\ \bibnamefont {Fan}}, \bibinfo
  {author} {\bibfnamefont {Lei}\ \bibnamefont {Wang}}, \ and\ \bibinfo {author}
  {\bibfnamefont {Pan}\ \bibnamefont {Zhang}},\ }\bibfield  {title} {\enquote
  {\bibinfo {title} {Unsupervised generative modeling using matrix product
  states},}\ }\href {\doibase 10.1103/PhysRevX.8.031012} {\bibfield  {journal}
  {\bibinfo  {journal} {Phys. Rev. X}\ }\textbf {\bibinfo {volume} {8}},\
  \bibinfo {pages} {031012} (\bibinfo {year} {2018})}\BibitemShut {NoStop}%
\bibitem [{\citenamefont {Glasser}\ \emph {et~al.}(2019)\citenamefont
  {Glasser}, \citenamefont {Sweke}, \citenamefont {Pancotti}, \citenamefont
  {Eisert},\ and\ \citenamefont {Cirac}}]{Glasser2019ExpressivePO}%
  \BibitemOpen
  \bibfield  {author} {\bibinfo {author} {\bibfnamefont {I.}~\bibnamefont
  {Glasser}}, \bibinfo {author} {\bibfnamefont {R.}~\bibnamefont {Sweke}},
  \bibinfo {author} {\bibfnamefont {Nicola}\ \bibnamefont {Pancotti}}, \bibinfo
  {author} {\bibfnamefont {J.}~\bibnamefont {Eisert}}, \ and\ \bibinfo {author}
  {\bibfnamefont {J.~I.}\ \bibnamefont {Cirac}},\ }\bibfield  {title} {\enquote
  {\bibinfo {title} {Expressive power of tensor-network factorizations for
  probabilistic modeling, with applications from hidden {Markov} models to
  quantum machine learning},}\ }in\ \href@noop {} {\emph {\bibinfo {booktitle}
  {NeurIPS}}}\ (\bibinfo {year} {2019})\BibitemShut {NoStop}%
\bibitem [{\citenamefont {Czarnik}\ and\ \citenamefont
  {Dziarmaga}(2014)}]{Czarnik2014}%
  \BibitemOpen
  \bibfield  {author} {\bibinfo {author} {\bibfnamefont {P.}~\bibnamefont
  {Czarnik}}\ and\ \bibinfo {author} {\bibfnamefont {J.}~\bibnamefont
  {Dziarmaga}},\ }\bibfield  {title} {\enquote {\bibinfo {title} {Fermionic
  projected entangled pair states at finite temperature},}\ }\href {\doibase
  10.1103/PhysRevB.90.035144} {\bibfield  {journal} {\bibinfo  {journal} {Phys.
  Rev. B}\ }\textbf {\bibinfo {volume} {90}},\ \bibinfo {pages} {035144}
  (\bibinfo {year} {2014})}\BibitemShut {NoStop}%
\bibitem [{Note1()}]{Note1}%
  \BibitemOpen
  \bibinfo {note} {In practice, we first employed a loss function without the
  denominator $1/ O^{\protect \rm sim}_\alpha $ in Figs.~\ref {Fig:Land}
  and~\ref {Fig:HAFC}, and then follows the exact form as Eq.~(\ref {Eq:Loss})
  in the cases of Figs.~\ref {Fig:CN} and \ref {Fig:TMGO}. Both schemes work
  well, and the design of the loss function have an empirical impact on its
  overall shape over the parameter space $\protect \mathcal {X}$, whose effects
  in the optimization efficiency will be carefully addressed in future
  studies.}\BibitemShut {Stop}%
\bibitem [{\citenamefont {LeCun}\ \emph {et~al.}(2015)\citenamefont {LeCun},
  \citenamefont {Bengio},\ and\ \citenamefont {Hinton}}]{LeCun2015}%
  \BibitemOpen
  \bibfield  {author} {\bibinfo {author} {\bibfnamefont {Yann}\ \bibnamefont
  {LeCun}}, \bibinfo {author} {\bibfnamefont {Yoshua}\ \bibnamefont {Bengio}},
  \ and\ \bibinfo {author} {\bibfnamefont {Geoffrey}\ \bibnamefont {Hinton}},\
  }\bibfield  {title} {\enquote {\bibinfo {title} {Deep learning},}\ }\href
  {\doibase 10.1038/nature14539} {\bibfield  {journal} {\bibinfo  {journal}
  {Nature}\ }\textbf {\bibinfo {volume} {521}},\ \bibinfo {pages} {436--444}
  (\bibinfo {year} {2015})}\BibitemShut {NoStop}%
\bibitem [{\citenamefont {{Shahriari}}\ \emph {et~al.}(2016)\citenamefont
  {{Shahriari}}, \citenamefont {{Swersky}}, \citenamefont {{Wang}},
  \citenamefont {{Adams}},\ and\ \citenamefont {{de Freitas}}}]{7352306}%
  \BibitemOpen
  \bibfield  {author} {\bibinfo {author} {\bibfnamefont {B.}~\bibnamefont
  {{Shahriari}}}, \bibinfo {author} {\bibfnamefont {K.}~\bibnamefont
  {{Swersky}}}, \bibinfo {author} {\bibfnamefont {Z.}~\bibnamefont {{Wang}}},
  \bibinfo {author} {\bibfnamefont {R.~P.}\ \bibnamefont {{Adams}}}, \ and\
  \bibinfo {author} {\bibfnamefont {N.}~\bibnamefont {{de Freitas}}},\
  }\bibfield  {title} {\enquote {\bibinfo {title} {Taking the human out of the
  loop: A review of {Bayesian} optimization},}\ }\href {\doibase
  10.1109/JPROC.2015.2494218} {\bibfield  {journal} {\bibinfo  {journal}
  {Proceedings of the IEEE}\ }\textbf {\bibinfo {volume} {104}},\ \bibinfo
  {pages} {148--175} (\bibinfo {year} {2016})}\BibitemShut {NoStop}%
\bibitem [{\citenamefont {Melnikov}\ \emph {et~al.}(2018)\citenamefont
  {Melnikov}, \citenamefont {Poulsen~Nautrup}, \citenamefont {Krenn},
  \citenamefont {Dunjko}, \citenamefont {Tiersch}, \citenamefont {Zeilinger},\
  and\ \citenamefont {Briegel}}]{Melnikov1221}%
  \BibitemOpen
  \bibfield  {author} {\bibinfo {author} {\bibfnamefont {Alexey~A.}\
  \bibnamefont {Melnikov}}, \bibinfo {author} {\bibfnamefont {Hendrik}\
  \bibnamefont {Poulsen~Nautrup}}, \bibinfo {author} {\bibfnamefont {Mario}\
  \bibnamefont {Krenn}}, \bibinfo {author} {\bibfnamefont {Vedran}\
  \bibnamefont {Dunjko}}, \bibinfo {author} {\bibfnamefont {Markus}\
  \bibnamefont {Tiersch}}, \bibinfo {author} {\bibfnamefont {Anton}\
  \bibnamefont {Zeilinger}}, \ and\ \bibinfo {author} {\bibfnamefont {Hans~J.}\
  \bibnamefont {Briegel}},\ }\bibfield  {title} {\enquote {\bibinfo {title}
  {Active learning machine learns to create new quantum experiments},}\ }\href
  {\doibase 10.1073/pnas.1714936115} {\bibfield  {journal} {\bibinfo  {journal}
  {Proceedings of the National Academy of Sciences}\ }\textbf {\bibinfo
  {volume} {115}},\ \bibinfo {pages} {1221--1226} (\bibinfo {year}
  {2018})}\BibitemShut {NoStop}%
\bibitem [{\citenamefont {van Tol}\ \emph {et~al.}(1971)\citenamefont {van
  Tol}, \citenamefont {Henkens},\ and\ \citenamefont {Poulis}}]{vanTol1971}%
  \BibitemOpen
  \bibfield  {author} {\bibinfo {author} {\bibfnamefont {M.~W.}\ \bibnamefont
  {van Tol}}, \bibinfo {author} {\bibfnamefont {L.~S. J.~M.}\ \bibnamefont
  {Henkens}}, \ and\ \bibinfo {author} {\bibfnamefont {N.~J.}\ \bibnamefont
  {Poulis}},\ }\bibfield  {title} {\enquote {\bibinfo {title} {High-field
  magnetic phase transition in
  $\mathrm{Cu}{(\mathrm{N}{\mathrm{O}}_{3})}_{2}\ifmmode\cdot\else\textperiodcentered\fi{}2\frac{1}{2}{{\mathrm{H}}}_{2}\mathrm{O}$},}\
  }\href {\doibase 10.1103/PhysRevLett.27.739} {\bibfield  {journal} {\bibinfo
  {journal} {Phys. Rev. Lett.}\ }\textbf {\bibinfo {volume} {27}},\ \bibinfo
  {pages} {739--741} (\bibinfo {year} {1971})}\BibitemShut {NoStop}%
\bibitem [{\citenamefont {Xu}\ \emph {et~al.}(2000)\citenamefont {Xu},
  \citenamefont {Broholm}, \citenamefont {Reich},\ and\ \citenamefont
  {Adams}}]{Xu2000}%
  \BibitemOpen
  \bibfield  {author} {\bibinfo {author} {\bibfnamefont {Guangyong}\
  \bibnamefont {Xu}}, \bibinfo {author} {\bibfnamefont {C.}~\bibnamefont
  {Broholm}}, \bibinfo {author} {\bibfnamefont {Daniel~H.}\ \bibnamefont
  {Reich}}, \ and\ \bibinfo {author} {\bibfnamefont {M.~A.}\ \bibnamefont
  {Adams}},\ }\bibfield  {title} {\enquote {\bibinfo {title} {Triplet waves in
  a quantum spin liquid},}\ }\href {\doibase 10.1103/PhysRevLett.84.4465}
  {\bibfield  {journal} {\bibinfo  {journal} {Phys. Rev. Lett.}\ }\textbf
  {\bibinfo {volume} {84}},\ \bibinfo {pages} {4465--4468} (\bibinfo {year}
  {2000})}\BibitemShut {NoStop}%
\bibitem [{\citenamefont {Xiang}\ \emph {et~al.}(2017)\citenamefont {Xiang},
  \citenamefont {Chen}, \citenamefont {Li}, \citenamefont {Sheng},
  \citenamefont {Su}, \citenamefont {Cheng}, \citenamefont {Chen},\ and\
  \citenamefont {Chen}}]{Xiang2017}%
  \BibitemOpen
  \bibfield  {author} {\bibinfo {author} {\bibfnamefont {Jun-Sen}\ \bibnamefont
  {Xiang}}, \bibinfo {author} {\bibfnamefont {Cong}\ \bibnamefont {Chen}},
  \bibinfo {author} {\bibfnamefont {Wei}\ \bibnamefont {Li}}, \bibinfo {author}
  {\bibfnamefont {Xian-Lei}\ \bibnamefont {Sheng}}, \bibinfo {author}
  {\bibfnamefont {Na}~\bibnamefont {Su}}, \bibinfo {author} {\bibfnamefont
  {Zhao-Hua}\ \bibnamefont {Cheng}}, \bibinfo {author} {\bibfnamefont {Qiang}\
  \bibnamefont {Chen}}, \ and\ \bibinfo {author} {\bibfnamefont {Zi-Yu}\
  \bibnamefont {Chen}},\ }\bibfield  {title} {\enquote {\bibinfo {title}
  {Criticality-enhanced magnetocaloric effect in quantum spin chain material
  copper nitrate},}\ }\href@noop {} {\bibfield  {journal} {\bibinfo  {journal}
  {Scientific Reports}\ }\textbf {\bibinfo {volume} {7}},\ \bibinfo {pages}
  {44643} (\bibinfo {year} {2017})}\BibitemShut {NoStop}%
\bibitem [{\citenamefont {Berger}\ \emph {et~al.}(1963)\citenamefont {Berger},
  \citenamefont {Friedberg},\ and\ \citenamefont {Schriempf}}]{Berger1963}%
  \BibitemOpen
  \bibfield  {author} {\bibinfo {author} {\bibfnamefont {L.}~\bibnamefont
  {Berger}}, \bibinfo {author} {\bibfnamefont {S.~A.}\ \bibnamefont
  {Friedberg}}, \ and\ \bibinfo {author} {\bibfnamefont {J.~T.}\ \bibnamefont
  {Schriempf}},\ }\bibfield  {title} {\enquote {\bibinfo {title} {Magnetic
  susceptibility of {Cu}
  ${(\mathrm{N}{\mathrm{O}}_{3})}_{2}$\ifmmode\cdot\else\textperiodcentered\fi{}2.5{${\mathrm{H}}_{2}${O}}
  at low temperature},}\ }\href {\doibase 10.1103/PhysRev.132.1057} {\bibfield
  {journal} {\bibinfo  {journal} {Phys. Rev.}\ }\textbf {\bibinfo {volume}
  {132}},\ \bibinfo {pages} {1057--1061} (\bibinfo {year} {1963})}\BibitemShut
  {NoStop}%
\bibitem [{\citenamefont {Li}\ \emph {et~al.}(2020{\natexlab{a}})\citenamefont
  {Li}, \citenamefont {Liao}, \citenamefont {Chen}, \citenamefont {Zeng},
  \citenamefont {Sheng}, \citenamefont {Qi}, \citenamefont {Meng},\ and\
  \citenamefont {Li}}]{Lih2020}%
  \BibitemOpen
  \bibfield  {author} {\bibinfo {author} {\bibfnamefont {Han}\ \bibnamefont
  {Li}}, \bibinfo {author} {\bibfnamefont {Yuan~Da}\ \bibnamefont {Liao}},
  \bibinfo {author} {\bibfnamefont {Bin-Bin}\ \bibnamefont {Chen}}, \bibinfo
  {author} {\bibfnamefont {Xu-Tao}\ \bibnamefont {Zeng}}, \bibinfo {author}
  {\bibfnamefont {Xian-Lei}\ \bibnamefont {Sheng}}, \bibinfo {author}
  {\bibfnamefont {Yang}\ \bibnamefont {Qi}}, \bibinfo {author} {\bibfnamefont
  {Zi~Yang}\ \bibnamefont {Meng}}, \ and\ \bibinfo {author} {\bibfnamefont
  {Wei}\ \bibnamefont {Li}},\ }\bibfield  {title} {\enquote {\bibinfo {title}
  {{Kosterlitz-Thouless} melting of magnetic order in the triangular quantum
  {Ising} material {TmMgGaO$_4$}},}\ }\href {\doibase
  10.1038/s41467-020-14907-8} {\bibfield  {journal} {\bibinfo  {journal} {Nat.
  Commun.}\ }\textbf {\bibinfo {volume} {11}},\ \bibinfo {pages} {1111}
  (\bibinfo {year} {2020}{\natexlab{a}})}\BibitemShut {NoStop}%
\bibitem [{\citenamefont {Li}\ \emph {et~al.}(2020{\natexlab{b}})\citenamefont
  {Li}, \citenamefont {Bachus}, \citenamefont {Deng}, \citenamefont {Schmidt},
  \citenamefont {Thoma}, \citenamefont {Hutanu}, \citenamefont {Tokiwa},
  \citenamefont {Tsirlin},\ and\ \citenamefont {Gegenwart}}]{Li2020}%
  \BibitemOpen
  \bibfield  {author} {\bibinfo {author} {\bibfnamefont {Y.}~\bibnamefont
  {Li}}, \bibinfo {author} {\bibfnamefont {S.}~\bibnamefont {Bachus}}, \bibinfo
  {author} {\bibfnamefont {H.}~\bibnamefont {Deng}}, \bibinfo {author}
  {\bibfnamefont {W.}~\bibnamefont {Schmidt}}, \bibinfo {author} {\bibfnamefont
  {H.}~\bibnamefont {Thoma}}, \bibinfo {author} {\bibfnamefont
  {V.}~\bibnamefont {Hutanu}}, \bibinfo {author} {\bibfnamefont
  {Y.}~\bibnamefont {Tokiwa}}, \bibinfo {author} {\bibfnamefont {A.~A.}\
  \bibnamefont {Tsirlin}}, \ and\ \bibinfo {author} {\bibfnamefont
  {P.}~\bibnamefont {Gegenwart}},\ }\bibfield  {title} {\enquote {\bibinfo
  {title} {Partial up-up-down order with the continuously distributed order
  parameter in the triangular antiferromagnet {${\mathrm{TmMgGaO}}_{4}$}},}\
  }\href {\doibase 10.1103/PhysRevX.10.011007} {\bibfield  {journal} {\bibinfo
  {journal} {Phys. Rev. X}\ }\textbf {\bibinfo {volume} {10}},\ \bibinfo
  {pages} {011007} (\bibinfo {year} {2020}{\natexlab{b}})}\BibitemShut
  {NoStop}%
\bibitem [{\citenamefont {Shen}\ \emph {et~al.}(2019)\citenamefont {Shen},
  \citenamefont {Liu}, \citenamefont {Qin}, \citenamefont {Shen}, \citenamefont
  {Li}, \citenamefont {Bewley}, \citenamefont {Schneidewind}, \citenamefont
  {Chen},\ and\ \citenamefont {Zhao}}]{Shen2019}%
  \BibitemOpen
  \bibfield  {author} {\bibinfo {author} {\bibfnamefont {Y.}~\bibnamefont
  {Shen}}, \bibinfo {author} {\bibfnamefont {C.}~\bibnamefont {Liu}}, \bibinfo
  {author} {\bibfnamefont {Y.}~\bibnamefont {Qin}}, \bibinfo {author}
  {\bibfnamefont {S.}~\bibnamefont {Shen}}, \bibinfo {author} {\bibfnamefont
  {Y.-D.}\ \bibnamefont {Li}}, \bibinfo {author} {\bibfnamefont
  {R.}~\bibnamefont {Bewley}}, \bibinfo {author} {\bibfnamefont
  {A.}~\bibnamefont {Schneidewind}}, \bibinfo {author} {\bibfnamefont
  {G.}~\bibnamefont {Chen}}, \ and\ \bibinfo {author} {\bibfnamefont
  {J.}~\bibnamefont {Zhao}},\ }\bibfield  {title} {\enquote {\bibinfo {title}
  {Intertwined dipolar and multipolar order in the triangular-lattice magnet
  {TmMgGaO$_4$}},}\ }\href {\doibase 10.1038/s41467-019-12410-3} {\bibfield
  {journal} {\bibinfo  {journal} {Nat. Commun.}\ }\textbf {\bibinfo {volume}
  {10}},\ \bibinfo {pages} {4530} (\bibinfo {year} {2019})}\BibitemShut
  {NoStop}%
\bibitem [{\citenamefont {Cevallos}\ \emph {et~al.}(2018)\citenamefont
  {Cevallos}, \citenamefont {Stolze}, \citenamefont {Kong},\ and\ \citenamefont
  {Cava}}]{Cava2018}%
  \BibitemOpen
  \bibfield  {author} {\bibinfo {author} {\bibfnamefont {F.~A.}\ \bibnamefont
  {Cevallos}}, \bibinfo {author} {\bibfnamefont {K.}~\bibnamefont {Stolze}},
  \bibinfo {author} {\bibfnamefont {T.}~\bibnamefont {Kong}}, \ and\ \bibinfo
  {author} {\bibfnamefont {R.~J.}\ \bibnamefont {Cava}},\ }\bibfield  {title}
  {\enquote {\bibinfo {title} {Anisotropic magnetic properties of the
  triangular plane lattice material {TmMgGaO}$_4$},}\ }\href {\doibase
  https://doi.org/10.1016/j.materresbull.2018.04.042} {\bibfield  {journal}
  {\bibinfo  {journal} {Mater. Res. Bull.}\ }\textbf {\bibinfo {volume}
  {105}},\ \bibinfo {pages} {154--158} (\bibinfo {year} {2018})}\BibitemShut
  {NoStop}%
\bibitem [{\citenamefont {Hu}\ \emph {et~al.}(2020)\citenamefont {Hu},
  \citenamefont {Ma}, \citenamefont {Liao}, \citenamefont {Li}, \citenamefont
  {Ma}, \citenamefont {Cui}, \citenamefont {Shangguan}, \citenamefont {Huang},
  \citenamefont {Qi}, \citenamefont {Li}, \citenamefont {Meng}, \citenamefont
  {Wen},\ and\ \citenamefont {Yu}}]{Hu2020}%
  \BibitemOpen
  \bibfield  {author} {\bibinfo {author} {\bibfnamefont {Ze}~\bibnamefont
  {Hu}}, \bibinfo {author} {\bibfnamefont {Zhen}\ \bibnamefont {Ma}}, \bibinfo
  {author} {\bibfnamefont {Yuan-Da}\ \bibnamefont {Liao}}, \bibinfo {author}
  {\bibfnamefont {Han}\ \bibnamefont {Li}}, \bibinfo {author} {\bibfnamefont
  {Chunsheng}\ \bibnamefont {Ma}}, \bibinfo {author} {\bibfnamefont
  {Yi}~\bibnamefont {Cui}}, \bibinfo {author} {\bibfnamefont {Yanyan}\
  \bibnamefont {Shangguan}}, \bibinfo {author} {\bibfnamefont {Zhentao}\
  \bibnamefont {Huang}}, \bibinfo {author} {\bibfnamefont {Yang}\ \bibnamefont
  {Qi}}, \bibinfo {author} {\bibfnamefont {Wei}\ \bibnamefont {Li}}, \bibinfo
  {author} {\bibfnamefont {Zi~Yang}\ \bibnamefont {Meng}}, \bibinfo {author}
  {\bibfnamefont {Jinsheng}\ \bibnamefont {Wen}}, \ and\ \bibinfo {author}
  {\bibfnamefont {Weiqiang}\ \bibnamefont {Yu}},\ }\bibfield  {title} {\enquote
  {\bibinfo {title} {Evidence of the {Berezinskii-Kosterlitz-Thouless} phase in
  a frustrated magnet},}\ }\href {\doibase 10.1038/s41467-020-19380-x}
  {\bibfield  {journal} {\bibinfo  {journal} {Nature Communications}\ }\textbf
  {\bibinfo {volume} {11}},\ \bibinfo {pages} {5631} (\bibinfo {year}
  {2020})}\BibitemShut {NoStop}%
\bibitem [{\citenamefont {Zhang}\ \emph {et~al.}(2020)\citenamefont {Zhang},
  \citenamefont {Li}, \citenamefont {Liu}, \citenamefont {Zhang}, \citenamefont
  {Ji}, \citenamefont {Jin}, \citenamefont {Chen}, \citenamefont {Wang},
  \citenamefont {Wang}, \citenamefont {Ma},\ and\ \citenamefont
  {Zhang}}]{Zhang2020effective}%
  \BibitemOpen
  \bibfield  {author} {\bibinfo {author} {\bibfnamefont {Zheng}\ \bibnamefont
  {Zhang}}, \bibinfo {author} {\bibfnamefont {Jianshu}\ \bibnamefont {Li}},
  \bibinfo {author} {\bibfnamefont {Weiwei}\ \bibnamefont {Liu}}, \bibinfo
  {author} {\bibfnamefont {Zhitao}\ \bibnamefont {Zhang}}, \bibinfo {author}
  {\bibfnamefont {Jianting}\ \bibnamefont {Ji}}, \bibinfo {author}
  {\bibfnamefont {Feng}\ \bibnamefont {Jin}}, \bibinfo {author} {\bibfnamefont
  {Rui}\ \bibnamefont {Chen}}, \bibinfo {author} {\bibfnamefont {Junfeng}\
  \bibnamefont {Wang}}, \bibinfo {author} {\bibfnamefont {Xiaoqun}\
  \bibnamefont {Wang}}, \bibinfo {author} {\bibfnamefont {Jie}\ \bibnamefont
  {Ma}}, \ and\ \bibinfo {author} {\bibfnamefont {Qingming}\ \bibnamefont
  {Zhang}},\ }\href@noop {} {\enquote {\bibinfo {title} {Effective magnetic
  {Hamiltonian} at finite temperatures for rare earth chalcogenides},}\ }
  (\bibinfo {year} {2020}),\ \Eprint {http://arxiv.org/abs/2011.06274}
  {arXiv:2011.06274 [cond-mat.str-el]} \BibitemShut {NoStop}%
\bibitem [{\citenamefont {Zhitomirsky}(2003)}]{Zhitomirsky2003}%
  \BibitemOpen
  \bibfield  {author} {\bibinfo {author} {\bibfnamefont {M.~E.}\ \bibnamefont
  {Zhitomirsky}},\ }\bibfield  {title} {\enquote {\bibinfo {title} {Enhanced
  magnetocaloric effect in frustrated magnets},}\ }\href {\doibase
  10.1103/PhysRevB.67.104421} {\bibfield  {journal} {\bibinfo  {journal} {Phys.
  Rev. B}\ }\textbf {\bibinfo {volume} {67}},\ \bibinfo {pages} {104421}
  (\bibinfo {year} {2003})}\BibitemShut {NoStop}%
\bibitem [{\citenamefont {Zhitomirsky}\ and\ \citenamefont
  {Honecker}(2004)}]{Zhitomirsky2004}%
  \BibitemOpen
  \bibfield  {author} {\bibinfo {author} {\bibfnamefont {M~E}\ \bibnamefont
  {Zhitomirsky}}\ and\ \bibinfo {author} {\bibfnamefont {A}~\bibnamefont
  {Honecker}},\ }\bibfield  {title} {\enquote {\bibinfo {title} {Magnetocaloric
  effect in one-dimensional antiferromagnets},}\ }\href {\doibase
  10.1088/1742-5468/2004/07/p07012} {\bibfield  {journal} {\bibinfo  {journal}
  {Journal of Statistical Mechanics: Theory and Experiment}\ }\textbf {\bibinfo
  {volume} {2004}},\ \bibinfo {pages} {P07012} (\bibinfo {year}
  {2004})}\BibitemShut {NoStop}%
\bibitem [{\citenamefont {Garst}\ and\ \citenamefont
  {Rosch}(2005)}]{Garst2005}%
  \BibitemOpen
  \bibfield  {author} {\bibinfo {author} {\bibfnamefont {Markus}\ \bibnamefont
  {Garst}}\ and\ \bibinfo {author} {\bibfnamefont {Achim}\ \bibnamefont
  {Rosch}},\ }\bibfield  {title} {\enquote {\bibinfo {title} {Sign change of
  the gr\"uneisen parameter and magnetocaloric effect near quantum critical
  points},}\ }\href {\doibase 10.1103/PhysRevB.72.205129} {\bibfield  {journal}
  {\bibinfo  {journal} {Phys. Rev. B}\ }\textbf {\bibinfo {volume} {72}},\
  \bibinfo {pages} {205129} (\bibinfo {year} {2005})}\BibitemShut {NoStop}%
\bibitem [{\citenamefont {Wolf}\ \emph {et~al.}(2011)\citenamefont {Wolf},
  \citenamefont {Tsui}, \citenamefont {Jaiswal-Nagar}, \citenamefont {Tutsch},
  \citenamefont {Honecker}, \citenamefont {Removi{\'c}-Langer}, \citenamefont
  {Hofmann}, \citenamefont {Prokofiev}, \citenamefont {Assmus}, \citenamefont
  {Donath},\ and\ \citenamefont {Lang}}]{Wolf2011}%
  \BibitemOpen
  \bibfield  {author} {\bibinfo {author} {\bibfnamefont {Bernd}\ \bibnamefont
  {Wolf}}, \bibinfo {author} {\bibfnamefont {Yeekin}\ \bibnamefont {Tsui}},
  \bibinfo {author} {\bibfnamefont {Deepshikha}\ \bibnamefont {Jaiswal-Nagar}},
  \bibinfo {author} {\bibfnamefont {Ulrich}\ \bibnamefont {Tutsch}}, \bibinfo
  {author} {\bibfnamefont {Andreas}\ \bibnamefont {Honecker}}, \bibinfo
  {author} {\bibfnamefont {Katarina}\ \bibnamefont {Removi{\'c}-Langer}},
  \bibinfo {author} {\bibfnamefont {Georg}\ \bibnamefont {Hofmann}}, \bibinfo
  {author} {\bibfnamefont {Andrey}\ \bibnamefont {Prokofiev}}, \bibinfo
  {author} {\bibfnamefont {Wolf}\ \bibnamefont {Assmus}}, \bibinfo {author}
  {\bibfnamefont {Guido}\ \bibnamefont {Donath}}, \ and\ \bibinfo {author}
  {\bibfnamefont {Michael}\ \bibnamefont {Lang}},\ }\bibfield  {title}
  {\enquote {\bibinfo {title} {Magnetocaloric effect and magnetic cooling near
  a field-induced quantum-critical point},}\ }\href {\doibase
  10.1073/pnas.1017047108} {\bibfield  {journal} {\bibinfo  {journal}
  {Proceedings of the National Academy of Sciences}\ }\textbf {\bibinfo
  {volume} {108}},\ \bibinfo {pages} {6862--6866} (\bibinfo {year}
  {2011})}\BibitemShut {NoStop}%
\bibitem [{\citenamefont {{Gegenwart}}(2016)}]{Gegenwart2016}%
  \BibitemOpen
  \bibfield  {author} {\bibinfo {author} {\bibfnamefont {Philipp}\ \bibnamefont
  {{Gegenwart}}},\ }\bibfield  {title} {\enquote {\bibinfo {title}
  {{Gr{\"u}neisen parameter studies on heavy fermion quantum criticality}},}\
  }\href {\doibase 10.1088/0034-4885/79/11/114502} {\bibfield  {journal}
  {\bibinfo  {journal} {Reports on Progress in Physics}\ }\textbf {\bibinfo
  {volume} {79}},\ \bibinfo {eid} {114502} (\bibinfo {year} {2016})},\ \Eprint
  {http://arxiv.org/abs/1608.04907} {arXiv:1608.04907 [cond-mat.str-el]}
  \BibitemShut {NoStop}%
\bibitem [{\citenamefont {Karbach}\ and\ \citenamefont
  {Stolze}(2005)}]{Karbach2005}%
  \BibitemOpen
  \bibfield  {author} {\bibinfo {author} {\bibfnamefont {Peter}\ \bibnamefont
  {Karbach}}\ and\ \bibinfo {author} {\bibfnamefont {Joachim}\ \bibnamefont
  {Stolze}},\ }\bibfield  {title} {\enquote {\bibinfo {title} {Spin chains as
  perfect quantum state mirrors},}\ }\href {\doibase
  10.1103/PhysRevA.72.030301} {\bibfield  {journal} {\bibinfo  {journal} {Phys.
  Rev. A}\ }\textbf {\bibinfo {volume} {72}},\ \bibinfo {pages} {030301}
  (\bibinfo {year} {2005})}\BibitemShut {NoStop}%
\bibitem [{\citenamefont {Cappellaro}\ \emph {et~al.}(2007)\citenamefont
  {Cappellaro}, \citenamefont {Ramanathan},\ and\ \citenamefont
  {Cory}}]{Cappellaro2007}%
  \BibitemOpen
  \bibfield  {author} {\bibinfo {author} {\bibfnamefont {P.}~\bibnamefont
  {Cappellaro}}, \bibinfo {author} {\bibfnamefont {C.}~\bibnamefont
  {Ramanathan}}, \ and\ \bibinfo {author} {\bibfnamefont {D.~G.}\ \bibnamefont
  {Cory}},\ }\bibfield  {title} {\enquote {\bibinfo {title} {Simulations of
  information transport in spin chains},}\ }\href {\doibase
  10.1103/PhysRevLett.99.250506} {\bibfield  {journal} {\bibinfo  {journal}
  {Phys. Rev. Lett.}\ }\textbf {\bibinfo {volume} {99}},\ \bibinfo {pages}
  {250506} (\bibinfo {year} {2007})}\BibitemShut {NoStop}%
\bibitem [{\citenamefont {Paszke}\ \emph {et~al.}(2019)\citenamefont {Paszke},
  \citenamefont {Gross}, \citenamefont {Massa}, \citenamefont {Lerer},
  \citenamefont {Bradbury}, \citenamefont {Chanan}, \citenamefont {Killeen},
  \citenamefont {Lin}, \citenamefont {Gimelshein}, \citenamefont {Antiga},
  \citenamefont {Desmaison}, \citenamefont {Kopf}, \citenamefont {Yang},
  \citenamefont {DeVito}, \citenamefont {Raison}, \citenamefont {Tejani},
  \citenamefont {Chilamkurthy}, \citenamefont {Steiner}, \citenamefont {Fang},
  \citenamefont {Bai},\ and\ \citenamefont {Chintala}}]{NEURIPS2019_9015}%
  \BibitemOpen
  \bibfield  {author} {\bibinfo {author} {\bibfnamefont {Adam}\ \bibnamefont
  {Paszke}}, \bibinfo {author} {\bibfnamefont {Sam}\ \bibnamefont {Gross}},
  \bibinfo {author} {\bibfnamefont {Francisco}\ \bibnamefont {Massa}}, \bibinfo
  {author} {\bibfnamefont {Adam}\ \bibnamefont {Lerer}}, \bibinfo {author}
  {\bibfnamefont {James}\ \bibnamefont {Bradbury}}, \bibinfo {author}
  {\bibfnamefont {Gregory}\ \bibnamefont {Chanan}}, \bibinfo {author}
  {\bibfnamefont {Trevor}\ \bibnamefont {Killeen}}, \bibinfo {author}
  {\bibfnamefont {Zeming}\ \bibnamefont {Lin}}, \bibinfo {author}
  {\bibfnamefont {Natalia}\ \bibnamefont {Gimelshein}}, \bibinfo {author}
  {\bibfnamefont {Luca}\ \bibnamefont {Antiga}}, \bibinfo {author}
  {\bibfnamefont {Alban}\ \bibnamefont {Desmaison}}, \bibinfo {author}
  {\bibfnamefont {Andreas}\ \bibnamefont {Kopf}}, \bibinfo {author}
  {\bibfnamefont {Edward}\ \bibnamefont {Yang}}, \bibinfo {author}
  {\bibfnamefont {Zachary}\ \bibnamefont {DeVito}}, \bibinfo {author}
  {\bibfnamefont {Martin}\ \bibnamefont {Raison}}, \bibinfo {author}
  {\bibfnamefont {Alykhan}\ \bibnamefont {Tejani}}, \bibinfo {author}
  {\bibfnamefont {Sasank}\ \bibnamefont {Chilamkurthy}}, \bibinfo {author}
  {\bibfnamefont {Benoit}\ \bibnamefont {Steiner}}, \bibinfo {author}
  {\bibfnamefont {Lu}~\bibnamefont {Fang}}, \bibinfo {author} {\bibfnamefont
  {Junjie}\ \bibnamefont {Bai}}, \ and\ \bibinfo {author} {\bibfnamefont
  {Soumith}\ \bibnamefont {Chintala}},\ }\bibfield  {title} {\enquote {\bibinfo
  {title} {Pytorch: An imperative style, high-performance deep learning
  library},}\ }in\ \href
  {http://papers.neurips.cc/paper/9015-pytorch-an-imperative-style-high-performance-deep-learning-library.pdf}
  {\emph {\bibinfo {booktitle} {Advances in Neural Information Processing
  Systems 32}}},\ \bibinfo {editor} {edited by\ \bibinfo {editor}
  {\bibfnamefont {H.}~\bibnamefont {Wallach}}, \bibinfo {editor} {\bibfnamefont
  {H.}~\bibnamefont {Larochelle}}, \bibinfo {editor} {\bibfnamefont
  {A.}~\bibnamefont {Beygelzimer}}, \bibinfo {editor} {\bibfnamefont
  {F.}~\bibnamefont {d\textquotesingle Alch\'{e}-Buc}}, \bibinfo {editor}
  {\bibfnamefont {E.}~\bibnamefont {Fox}}, \ and\ \bibinfo {editor}
  {\bibfnamefont {R.}~\bibnamefont {Garnett}}}\ (\bibinfo  {publisher} {Curran
  Associates, Inc.},\ \bibinfo {year} {2019})\ pp.\ \bibinfo {pages}
  {8024--8035}\BibitemShut {NoStop}%
\bibitem [{\citenamefont {Lizotte}(2008)}]{lizotte2008}%
  \BibitemOpen
  \bibfield  {author} {\bibinfo {author} {\bibfnamefont {Daniel~James}\
  \bibnamefont {Lizotte}},\ }\emph {\bibinfo {title} {Practical Bayesian
  Optimization}},\ \href@noop {} {Ph.D. thesis},\ \bibinfo {address} {CAN}
  (\bibinfo {year} {2008}),\ \bibinfo {note} {aAINR46365}\BibitemShut {NoStop}%
\bibitem [{\citenamefont {Nogueira}(2014--)}]{BOpackage}%
  \BibitemOpen
  \bibfield  {author} {\bibinfo {author} {\bibfnamefont {Fernando}\
  \bibnamefont {Nogueira}},\ }\href
  {https://github.com/fmfn/BayesianOptimization} {\enquote {\bibinfo {title}
  {{Bayesian Optimization}: Open source constrained global optimization tool
  for {Python}},}\ } (\bibinfo {year} {2014--})\BibitemShut {NoStop}%
\end{thebibliography}%
%
\newpage
\clearpage
\onecolumngrid
\mbox{}
\begin{center}
\textbf{\large Supplementary Materials: \\Learning Effective Spin Hamiltonian of Quantum Magnet}\\
Yu \textit{et al}.
\end{center}

\date{\today}

\setcounter{section}{0}
\setcounter{figure}{0}
\setcounter{equation}{0}
\renewcommand{\thesection}{\Alph{section}}
\renewcommand{\theequation}{S\arabic{equation}}
\renewcommand{\thefigure}{S\arabic{figure}}


\section{Automatic Hamiltonian Searching Algorithms}\label{SMSec:AHS}
Below we list three algorithms adopted in 
our Hamiltonian searching, which include 
the random grid (Algorithm~\ref{Alg:RGS}),
auto-gradient (Algorithm~\ref{Alg:BFGS}), 
and the Bayesian (Algorithm~\ref{Alg:BO}) methods. 
These three searching schemes can be 
combined with various many-body
thermodynamics solvers in a very flexible manner, 
rendering different resolutions 
in determining the Hamiltonian parameters.\\

\begin{algorithm2e}[H]
Discretize the parameter space into a uniform grid with n nodes ($\mathbf{x}_1, ..., \mathbf{x}_n$)\;
\For{$i = 1$ \KwTo $n$}{ Random select one of the unevaluated nodes $\mathbf{x}_i$ 
and calculate the $\mathcal{L}(\mathbf{x}_i)$\; Mark $\mathbf{x}_i$ as evaluated.
}
\caption{Random Grid Searching}
\label{Alg:RGS}
\end{algorithm2e}

\begin{algorithm2e}[H]
\For{$i = 1$ \KwTo $n$}{
 Random choose a starting point $\mathbf{x_i}$\;
 \For{$j = 1$ \KwTo $n$}{
 	$x_{i, j+1} = x_{i, j} + \lambda B^{-1} \nabla f(x_{i, j})$, where $B$ is an approximate Hessian\;
 	\If {EOF}{
 	 go to 2
 	}
 }
 }
\caption{Multi-Restart Auto-Gradient}
\label{Alg:BFGS}
\end{algorithm2e} 

\begin{algorithm2e}[H]
 Initialize a statistical model\;
 \For{$i = 1$ \KwTo $n$}{
  select the next point $\mathbf{x}_{i+1}$ to evaluate by maximizing the 
  acquisition function $\mathbf{x}_{i+1} = \mathop{\arg\max}_{\mathbf{x}} \alpha(\mathbf{x}; \mathcal{D}_i)$\;
  evaluate objective function $y_{i+1}$ at $\mathbf{x}_{i+1}$\;
  Augment data $\mathcal{D}_{i+1} = \{\mathcal{D}_i, (\mathbf{x}_{i+1}), y_{i+1}\}$\;
  update statistical model with $\mathcal{D}_{i+1}$
 }
    \caption{Bayesian Optimization}
\label{Alg:BO}
\end{algorithm2e} 

\section{Automatic Differentiation}\label{SMSec:AD}
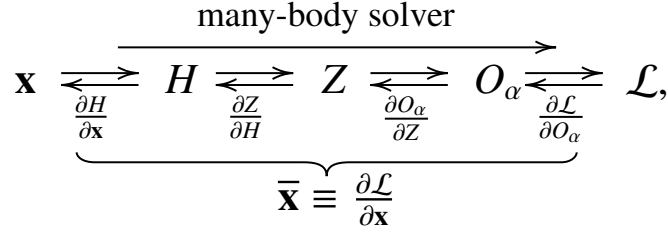
\begin{figure}[!tbp]
\tikzset{every picture/.style={scale=1.5,line width=0.75pt,inner sep=0.75pt,anchor=north west}} 
\newcommand\arrowhead{(8,-3)..controls(5,-1)and(3,-0.2)..(0,0)..controls(3,0.2)and(5,1)..(8,3)}
\newcommand\UnderBrace{(58,88)..controls(58,90)and(60,93)..(65,93)--(146,93)..controls(152,93)and(155,95)..(155,100)..controls(155,95)and (158,93)..(164,93)--(246,93)..controls(250,93)and(252,90)..(252,88)}  
\begin{tikzpicture}[x=.65pt,y=0.5pt,yscale=-1,xscale=1]
\path (100,0); 
\draw  (75,38)   -- (243,38);\draw [shift={(245,38)}, rotate = 180] \arrowhead;
\draw  (53,51)   --   (81,51);\draw [shift={(83,51)}, rotate = 180]\arrowhead;
\draw  (83,57)   --   (55,57);\draw [shift={(53,57)}, rotate = 360]\arrowhead;
\draw  (113,51) -- (141,51);\draw [shift={(143,51)}, rotate = 180]\arrowhead;
\draw  (143,57) -- (115,57);\draw [shift={(113,57)}, rotate = 360] \arrowhead;
\draw  (173,51) -- (201,51);\draw [shift={(203,51)}, rotate = 180] \arrowhead;
\draw  (203,57) -- (175,57);\draw [shift={(173,57)}, rotate = 360]\arrowhead;
\draw  (233,51) -- (261,51);\draw [shift={(263,51)}, rotate = 180]\arrowhead;
\draw  (263,57) -- (235,57);\draw [shift={(233,57)}, rotate = 360] \arrowhead;
\draw  [shift={(1,3)}]\UnderBrace;
\draw (104,13) node [scale=1.5] {{\fontfamily{ptm}\selectfont many-body solver}};
\draw (33,48)   node [scale=1.8]  {$\mathbf{x}$};
\draw (91,44)   node [scale=1.8]  {$H$};
\draw (152,44) node [scale=1.8]  {$Z$};
\draw (211,44) node [scale=1.8]  {$O_\alpha$};
\draw (270,44) node [scale=1.8]  {$\mathcal{L}$,};
\draw (55,60)  node  [scale=1.5]  {$\tfrac{\partial H}{\partial \mathbf{x}}$};
\draw (115,60) node [scale=1.5]  {$\tfrac{\partial Z}{\partial H}$};
\draw (175,60) node [scale=1.5]  {$\tfrac{\partial O_\alpha}{\partial Z}$};
\draw (235,60) node [scale=1.5]  {$\tfrac{\partial \mathcal{L}}{\partial O_\alpha}$};
\draw (135,99) node [scale=1.8]  {$\overline{\mathbf{x}} \equiv \tfrac{\partial \mathcal{L}}{\partial \mathbf{x}}$};
\end{tikzpicture}
\caption{A typical computational graph of the quantum 
many-body calculations, with the forward process indicated 
by all the right-directed lines, and the backward process 
by the left-directed lines.}
\label{Fig:CompGraph}
\end{figure}

In this section, we provide more details of automatic 
differentiation used in our auto-gradient scheme.
Automatic differentiation is a well-developed technique 
in neural networks and deep learning~\cite{LeCun2015}. 
A central ingredient of automatic differentiation is the 
so-called computational graph (see \Fig{Fig:CompGraph} 
for a typical computational graph for many-body calculations). 
To generate such a computational graph, one starts with the input parameters, goes through a number of intermediate computation nodes, and ends up with the final loss function.

To be specific, for the quantum many-body problems afore-mentioned in main text, starting with several Hamiltonian parameters, e.g., ${\bf x}\equiv\{J, \Delta, g, \cdots\}$, 
one defines the many-body model Hamiltonian $H({\bf x})$.
Given it either ED or thermal tensor network calculations, the partition function $Z$ and thereafter thermodynamic observables $\{O_\alpha\}$, 
can be obtained. Basing on the calculated observables $\{O_\alpha\}$, a loss function can be properly designed 
[cf. \Eq{Eq:Loss}]. 
The above procedure constitutes a forward evaluation 
of the loss function, and henceforth a computational graph 
${\bf x} \to H \to Z \to O_\alpha \to \mathcal{L}$ is generated 
(cf. the right-directed lines in \Fig{Fig:CompGraph}).

On the fly of the forward process, the derivatives between adjacent computation nodes, i.e. 
$\{\frac{\partial H}{\partial\bf x},  \frac{\partial Z}{\partial H}, \frac{\partial O_\alpha}{\partial Z}, \frac{\partial\mathcal{L}}{\partial O_\alpha}\}$, are stored. 
Thus the derivatives of loss function with respect to the input parameters can be evaluated automatically via a chain rule, 
\begin{equation}\label{Eq:forward}
\frac{\partial\mathcal{L}}{\partial\bf x} 
= \frac{\partial\mathcal{L}}{\partial O_\alpha}
\,\frac{\partial O_\alpha}{\partial Z} 
\,\frac{\partial Z}{\partial H} 
\,\frac{\partial H}{\partial\bf x}.
\end{equation}
In our cases, since the number of input parameters 
[components in $\bf{x}$, typically a few to $O(10)$]
is larger than the output (just a single value of 
loss $\mathcal{L}$), it is therefore more efficient to evaluate 
\Eq{Eq:forward} following the reverse-mode 
automatic differentiation (i.e., from left to right 
on the right-hand side of the equation). 
In this work, we have implemented a 
differentiable ED calculation with Pytorch \cite{NEURIPS2019_9015}, and the generalization to 
tensor networks is also feasible~\cite{Liao2019,Chen2020}.

\begin{figure}[!tbp]
\includegraphics[angle=0,width=0.9\linewidth]{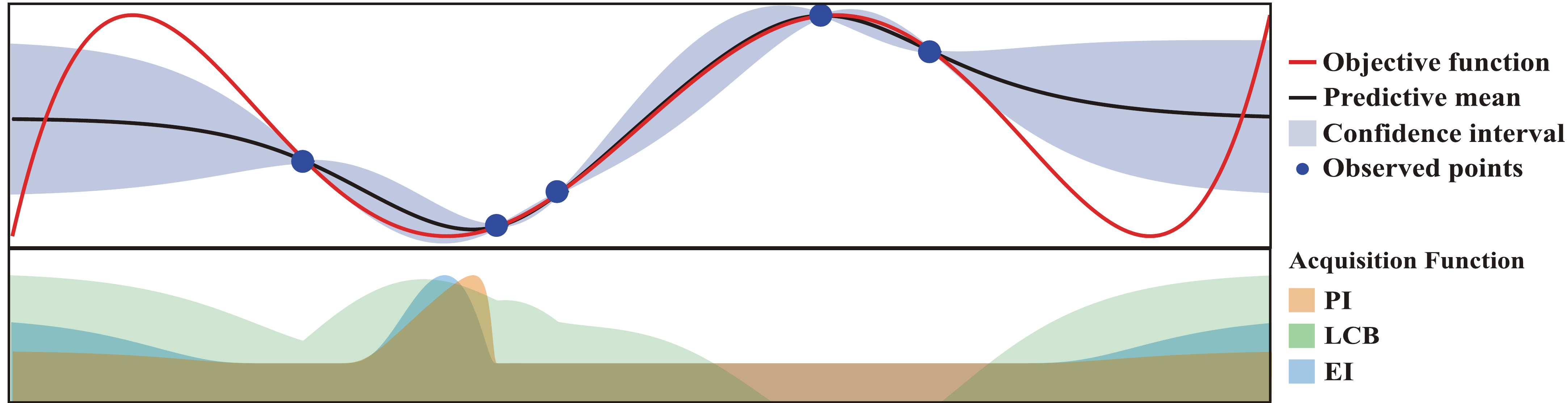}
\caption{Red line in the upper panel represent the unknown objective function, with blue dots the evaluated
points. The black line and shade represent the predicted mean and confidence interval, respectively. 
Various shades in the lower panel corespondent to different acquisition functions.}
\label{Fig:BO}
\end{figure}

\section{Bayesian Optimization with Gaussian Process: Kernel Function, ARD and Acquisition Function}\label{SMSec:BO}
As shown in Algorithm \ref{Alg:BO} and Fig.~\ref{Fig:BO}, in the Bayesian optimization we need to iteratively update a statistical model that can be used to estimate the overall landscape of $\mathcal{L}$, based on history queries. In this work, we choose a commonly used statistical model called Gaussian process, which fits well our problem and is denoted as 
\begin{equation}
\mathcal{GP}: \mathcal{X}, \mathcal{D} 
\longrightarrow \mu, \sigma
\end{equation}
 where $\mathcal{X}$ is the parameter space spanned by the parameter vectors $\bf{x}$, 
 which could include, in practice, components $J, g, \Delta$, etc. The set of total 
 $n$ history queries is noted as $\mathcal{D}_n = (({\bf{x}}_1, y_1), ({\bf{x}}_2, y_2), ..., ({\bf{x}}_n, y_n))$, 
 with $y_i$ being the evaluated function value at parameter $\bf{x}_i$, i.e., 
 $\mathcal{L}(\bf{x}_i)$. Then by assuming a joint multivariate 
 Gaussian distribution over $(y_1, y_2, ..., y_n; y_{n+1})$, 
 with the covariances characterized by a kernel function 
 $k(\bf{x}_i, \bf{x}_j)$, and $y_{n+1}$ to be estimated 
 at ${\bf{x}}_{n+1}$, we can compute a \textit{posterior} 
 distribution of $y_{n+1} \sim \mathcal{N}(\mu_n, \sigma_n^2)$ by
\begin{align}
    \mu_n(\mathbf{x}_{n+1}) & = \mathbf{k}(\mathbf{x}_{n+1})\tran \mathbf{K}^{-1} \mathbf{y}, \\
    \sigma_n^2(\mathbf{x}_{n+1}) & = \mathit{k}(\mathbf{x}_{n+1}, \mathbf{x}_{n+1}) 
    - \mathbf{k}(\mathbf{x}_{n+1})\tran \mathbf{K}^{-1}  \mathbf{k}(\mathbf{x}_{n+1}),
\end{align}
where a constant zero prior mean is assumed in the space $\mathcal{X}$.
${\bf{y}}=(y_1, y_2, ..., y_n)\tran$ is the vector of evaluated function values,
$\mathbf{k}(\mathbf{x})\tran =\left (k(\mathbf{x}, \mathbf{x_1}), k(\mathbf{x}, \mathbf{x_2}), ...\right)$
and $\mathbf{K}_{i,j} = k(\mathbf{x}_i,\mathbf{x}_j)$
are respectively the covariance vector and matrix , 
where $\mathbf{x_1}, \mathbf{x_2}, ..., \mathbf{x_n}$ 
represent the calculated parameter points in the history queries. 
The quality of GP regression to fit the real landscape is determined 
by the choice of kernel function $k(\bf{x}, \bf{x}')$. 
In practice, we chose a Mat\'{e}rn$-5$ kernel, i.e., 
\begin{align}
	\mathllap{\mathit{k}_{Mat\acute{e}rn5}(\bf{x}, \bf{x}')} &= \theta_0^2 \exp(-\sqrt{5})(1 + \sqrt{5}r + \frac{5}{3} r^2),
\end{align}
where in the kernel function $r^2=(\mathbf{x} - \mathbf{x}')\tran\mathbf{\Lambda} (\mathbf{x} - \mathbf{x}')$ 
and $\mathbf{\Lambda}$ is a 
diagonal matrix with length scale $\theta_i^2$. 
Then we are left with hyperparameters 
$\theta_i$ to be determined,
which describe the scale of the 
kernel function for each parameter. 
Fortunately, the GP model provide 
us a nice analytical expression 
of for the marginal likelihood 
with the following expression,
\begin{equation}
\log p(\mathbf{y} | \mathbf{x}, \theta) = -\frac{1}{2} \mathbf{y}\tran (\mathbf{K}^\theta)^{-1} \mathbf{y} - \frac{1}{2} \log |\mathbf{K}^\theta| - \frac{n}{2}\log(2\pi).
\end{equation}
Note here $\theta$ represents a set 
of all the hyperparameters, 
and we can easily compute $\theta^*$ 
that maximize the marginal likelihood, 
as long as the kernel is differentiable with respect to $\theta$. 
By denoting $\theta^* = \theta_{\rm ML}$, 
we take it as a point estimator for our hyperparameters. 
Besides, one can also use a maximum 
a posteriori estimation $\theta_{\rm MAP}$ 
as the kernel parameters. This technique 
is often referred to as automatic 
relevance determination (ARD) kernels.

With the estimated mean $\mu_n$ and variance $\sigma_n$, we can estimate the landscape $\mathcal{L}(\mathbf{x})$, and choose the next point $\mathbf{x}_{n+1}$ by maximizing an acquisition function $\alpha(\mathbf{x})$, i.e., $\mathbf{x}_{n+1}  = \arg \rm \max_\mathbf{x} \alpha(\mathbf{x})$. A careful design of acquisition function is needed to balance the efficiency and exploration of the parameter space. Here we introduce three very popular acquisition functions that are commonly adopted: probability of improvement (PI), expected improvement (EI) and lower confidence bound (LCB). To be clear of the notations, the term ``improvement'' in the context of minimization means the diminution of the minimum. The three acquisition functions are
\begin{align}
    &\begin{aligned}
        \mathllap{\alpha_{\rm PI}(\mathbf{x}; \mathcal{D}_n)} &= \mathbb{P}[\mathcal{L}(\mathbf{x}) \leq \tau ] = \mathit{\Phi}\left( - \frac{\mu_n(\mathbf{x})-\tau}{\sigma_n(\mathbf{x})}\right),
    \end{aligned}\\
    &\begin{aligned}
        \mathllap{\alpha_{\rm EI}(\mathbf{x}; \mathcal{D}_n)} &= \mathbb{E}[\tau - \mathcal{L}(\mathbf{x})] = (\tau - \mu_n(\mathbf{x})) \mathit{\Phi}\left(\frac{\tau - \mu_n(\mathbf{x})}{\sigma_n(\mathbf{x})}\right)  + \sigma_n(\mathbf{x})\phi\left(\frac{\tau - \mu_n(\mathbf{x})}{\sigma_n(\mathbf{x})}\right),
    \end{aligned}\\
    &\begin{aligned}
        \mathllap{\alpha_{\rm LCB}(\mathbf{x}; \mathcal{D}_n)} 
        = \mu_n(\mathbf{x}) - \kappa \sigma_n(\mathbf{x}) ,
    \end{aligned}
\end{align}
where where $\phi$ and $\Phi$ denote the PDF and CDF of normal distribution, and $\tau = \mathcal{L}_{\rm min} -  \xi$ with $\xi$ an adjustable empirical parameter, and so is $\kappa$ in LCB. It has been shown in previous works that $\xi = 0.01\sigma_f$, with $\sigma_f$ being the standard deviation of $\bf{y}$, constitutes a setting that has an overall very good performance~\cite{lizotte2008}, which is adopted in this work. A visualization of Gaussian process and acquisition functions is in Fig.~\ref{Fig:BO}. An open-source python package was used in this work for numerical experiments \cite{BOpackage}. Moreover, one could also choose information-based acquisition function or a portfolio of acquisition strategies to balance the efficiency and over-all exploration.

\begin{figure}[tbp]
\includegraphics[angle=0,width=0.4\linewidth]{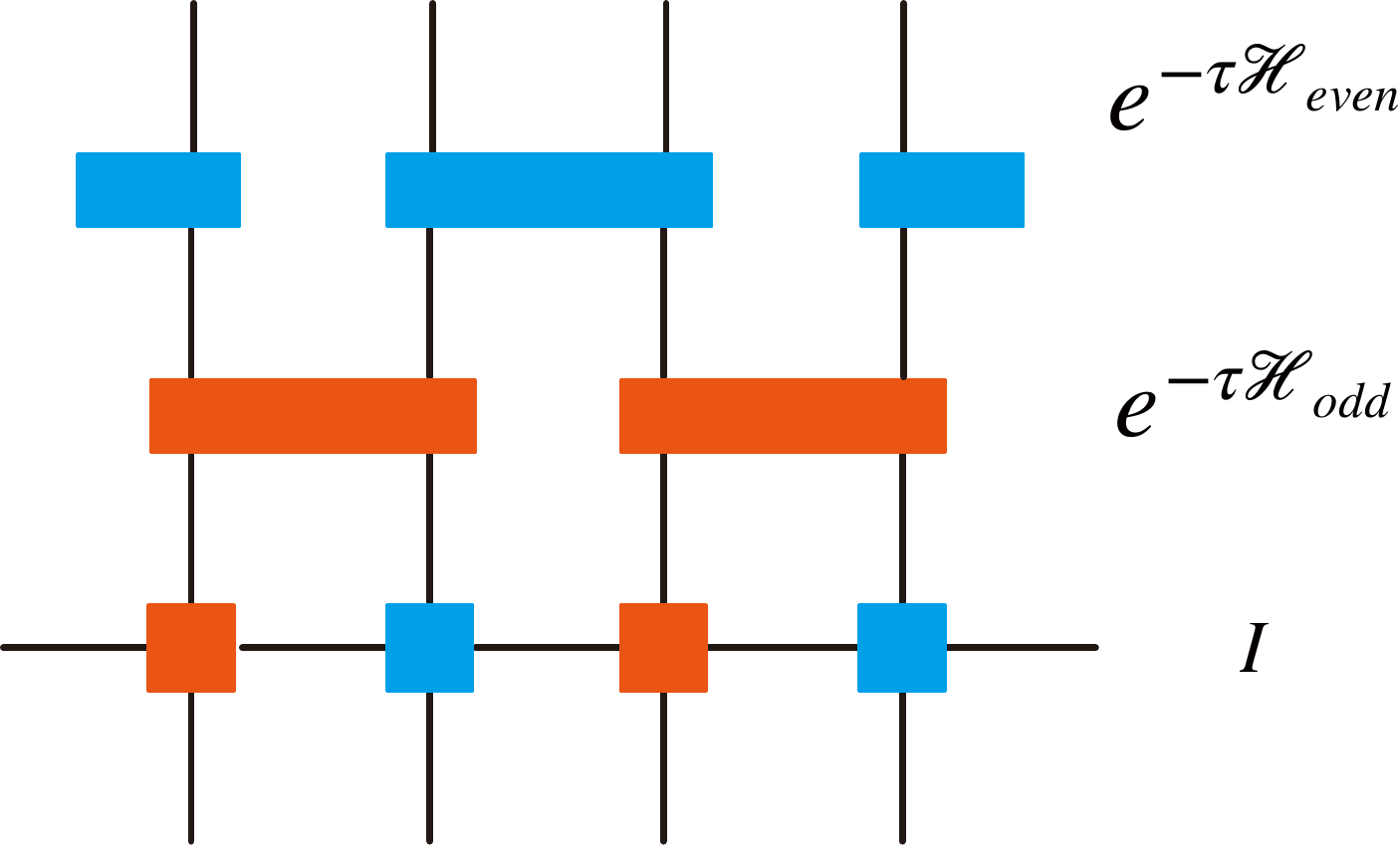}
\caption{Tensor network representation 
of the density matrix $\rho(\tau)$ [cf. \Eq{eq:TSD}]. 
The bottom line is the infinite-temperature density 
operator $\rho(0)=I$. The above blue/red blocks 
indicate the Trotter gates on even/odd bonds.}
\label{Fig:LTRG}
\end{figure}

\section{Quantum Many-body Calculation Methods}
\label{SMSec:QMBC}

In this section, we introduce the basic idea 
of some quantum many-body calculation methods, 
including exact diagonalization (ED) and linearized tensor renormalization group algorithm (LTRG)~\cite{Li2011,Dong2017}. 
To calculate the thermodynamic properties of a quantum many-body systems, one needs to obtain 
the partition function $\mathcal{Z}=\mathrm{tr}(\hat\rho)=\mathrm{tr}(e^{-\beta \mathcal{H}})$ 
with high precision. For quantum lattice models with $d$-dimension local Hilbert space ($d=2$ for spin-1/2 systems), 
the $\mathcal{N}$-site many-body basis totally takes a $d^\mathcal{N}$-dimension space, and rendering the Hamiltonian 
$\mathcal{H}$ being a $d^\mathcal{N}\times d^\mathcal{N}$ matrix. 

Limited by the numerical resources, currently one can only store and diagonalize a spin-1/2 Hamiltonian with size of $\mathcal{N}\lesssim20$ sites.
For those small systems, we diagonalize $\mathcal{H}$ by an invertible matrix $\mathcal{U}$ as
\begin{equation}
	\mathcal{H} = \mathcal{U} \mathcal{D} \mathcal{U}^{-1}
	\label{ED}
\end{equation}
with $\mathcal{D}$ diagonalized. Thereafter, we obtain the density matrix of the system at the inverse temperature $\beta$
\begin{equation}
	\rho = e^{-\beta \mathcal{H}} = \mathcal{U}\, e^{-\beta \mathcal{D}}\, \mathcal{U}^{-1},
\end{equation}
the partition function
\begin{equation}
	\mathcal{Z} = \text{tr} (e^{-\beta \mathcal{H}}) = \text{tr} (e^{-\beta \mathcal{D}}),
	\label{eq:par}
\end{equation}
and thus other thermodynamic quantities.

\begin{figure}[!tbp]
\includegraphics[angle=0,width=1\linewidth]{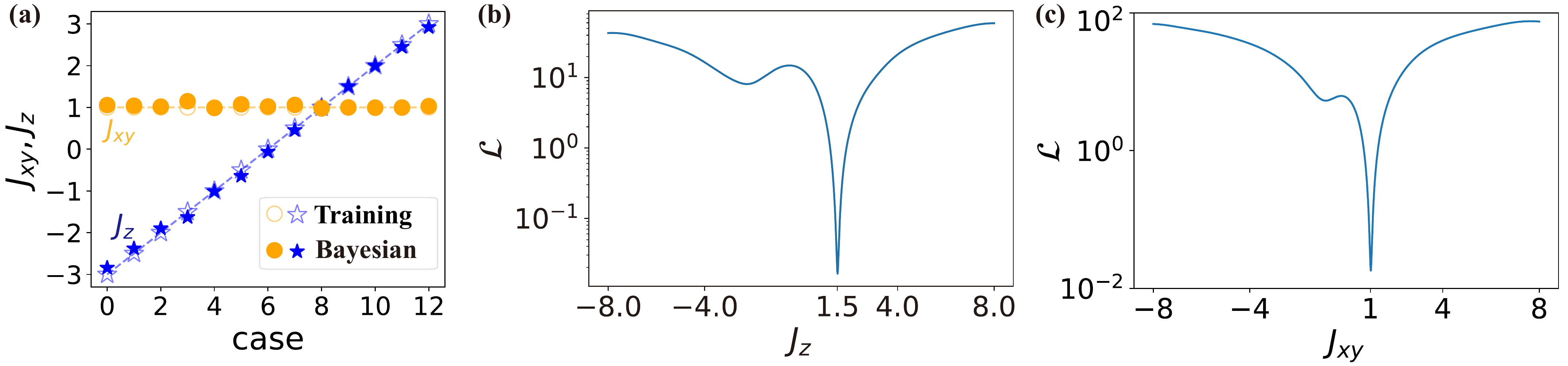}
\caption{(a) The optimal parameter found after 150 Bayesian optimization iterations. Dashed line indicate the training data, the stars and dots mark respectively the determined parameters $J_z$ and $J_{xy}$. (b) A cross cut of the 2D landscape $\mathcal{L}$ in Fig.~\ref{Fig:Land} of the main text, $\mathcal{L}(1.025, J_{z})$ vs. $J_z$, with $J_{xy} = 1.025$ fixed at the predicted optimal value. (c) shows the cross cut of the 2D landscape $\mathcal{L}$ in Fig.~\ref{Fig:Land}, $\mathcal{L}(J_{xy}, 1.49)$ vs. $J_{xy}$, with a fixed $J_{z} = 1.49$ at its predicted optimal value. A sharp deep near the optimal parameter point can be clearly observed in both (b) and (c) panels.}
\label{Fig:XXZ}
\end{figure}

For larger system sizes, we resort to thermal tensor network methods, to be specific, LTRG in this work. 
The basic idea of LTRG is to, firstly slice the lower-temperature density matrix $\rho(\beta)=e^{-\beta\mathcal{H}}$ 
into $N$ small slots $\tau=\beta/N$, i.e.
\begin{equation}
	\rho(\beta) = e^{-\beta \mathcal{H}} =  (e^{-\tau \mathcal{H}})^N.
	\label{eq:DM}
\end{equation}
For a one-dimensional system that contains only the nearest-neighboring interactions, the Hamiltonian can be divided into odd and even parts such that
\begin{equation}
	\mathcal{H} = \mathcal{H}_{odd} + \mathcal{H}_{even}.
\end{equation}
Generally, these two parts are non-commutative, so we need to use Trotter-Suzuki decomposition to separate the two terms as 
\begin{equation}
	\rho(\tau) = e^{-\tau (\mathcal{H}_{odd}+\mathcal{H}_{even})} = e^{-\tau \mathcal{H}_{odd}}e^{-\tau \mathcal{H}_{even}} + O(\tau^2).
	\label{eq:TSD}
\end{equation}
Now we arrive at 
\begin{equation}
	\rho(\beta) = [\rho(\tau)]^N \simeq (e^{-\tau \mathcal{H}_{even}} e^{-\tau \mathcal{H}_{odd}})^N
	\label{eq:DM1}
\end{equation}
with discretization error $O(\tau^2)$. The tensor network representation of $\rho(\tau)$ is shown in \Fig{Fig:LTRG}, 
with an infinite-temperature density matrix (identity matrix) $\rho(0)=I$ explicitly shown. 
Therefore, \Eq{eq:DM1} can be viewed as a cooling process following a linear temperature gird, 
i.e. $0\to\tau\to2\tau\to3\tau\to\cdots\to N\tau\equiv\beta$. 
The partition function $\mathcal{Z}=\mathrm{tr}[\rho(\beta)]$ can thus be obtained by fully contracting the tensor network \Eq{eq:DM1}, 
and the relevant thermodynamic quantities can be obtained directly from the partition function as, 
\begin{gather}
	f =-\frac{1}{\beta} \text{ln}\mathcal{Z} \\
	C = \beta^2 \frac{\partial^2 \text{ln} \mathcal{Z}}{\partial \beta^2}\\
	M = -\frac{\partial f}{\partial h}\\
	\chi = \frac{M}{h}
\end{gather}
where $f$ is the free energy, $C$ is the heat capacity, $h$ is the magnetic field strength, 
$M$ is the magnetization, and $\chi$ is the magnetic susceptibility.

\begin{figure}[!tbp]
\includegraphics[angle=0,width=1\linewidth]{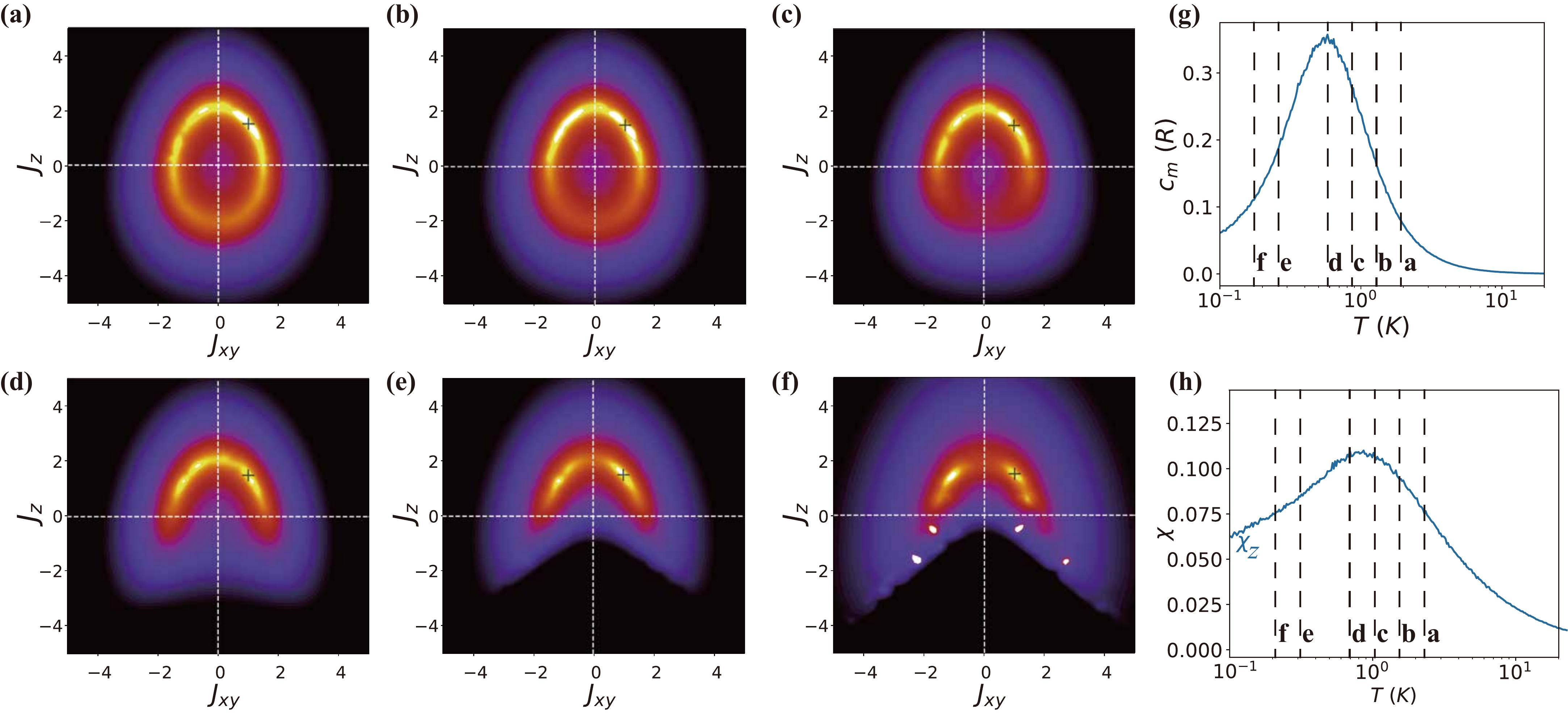}
\caption{(a-f) Landscape $\mathcal{L}$ interpolated by a grid search of a $30 \times 30$ grid with varying cut-off temperature $T_{\rm cut}$, as shown in (g, h). Inconsistent points in (f) are due to interpolation errors. (g, h) indicate the fitted thermodynamic quantities and various $T_{\rm cut}$ values, above which the thermodynamics data are used for fitting.}
\label{Fig:SmTcut1chi}
\end{figure}

\begin{figure}[!tbp]
\includegraphics[angle=0,width=1\linewidth]{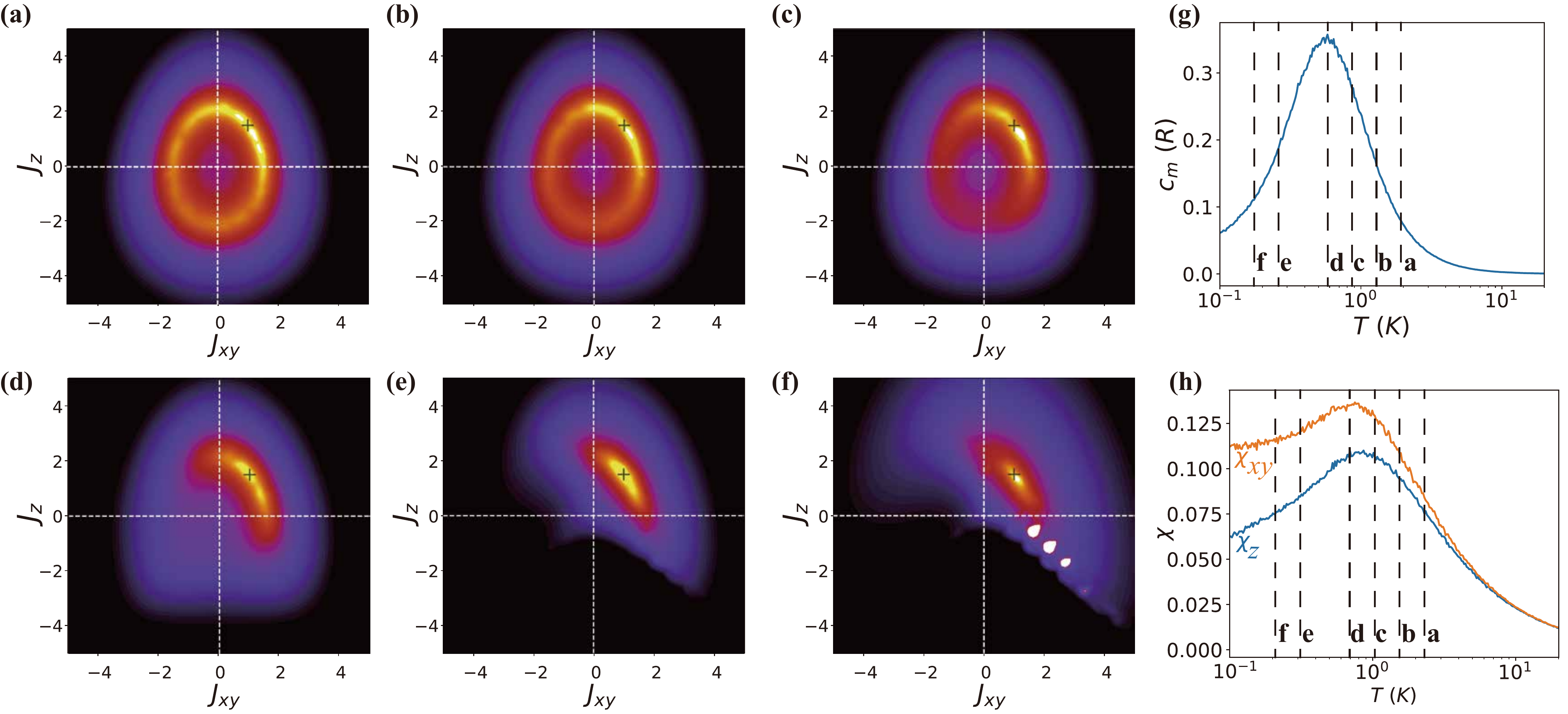}
\caption{Same layout as Fig.~\ref{Fig:SmTcut1chi}, with $\chi_{xy}$ included in the automatic parameter searching. }
\label{Fig:SmTcut2chi}
\end{figure}

\section{More Results on the XXZ HAFC systems}
\label{SMSec:XXZ}
With the training data from the given XXZ spin-chain model, here we show more cases to further validate the robustness of our method. For clarity, the ED calculation of 10 sites XXZ spin chain is used as a rudimentary many-body solver, although in practice we find ED calculations with 8-12 sites lead to virtually the same performance. In Fig.~\ref{Fig:XXZ}, we choose different $J_z$ $\in [-3, 3]$ and a fixed $J_{xy}=1$, and find that the Bayesian optimization can always refind the correct parameters in all cases and thus constitutes a robust approach.

Then we show the landscapes obtained at different $T_{\rm cut}$ temperatures, fitting jointly the specific heat $C_m$ and susceptibility $\chi_z$, and observed various landscapes in Fig.~\ref{Fig:SmTcut1chi}. We observe a symmetric landscape on $J_{xy}$, due to the identical energy spectra, as well as thermodynamics $C_m$ and $\chi_{xy}$, for two models with $\pm |J_{xy}|$. By introducing the in-plane susceptibility $\chi_{xy}$, we can lift this degeneracy, as shown in Fig.~\ref{Fig:SmTcut2chi}.

Notably, we find that for a rather high $T_{\rm cut}$, an oval ring with $J_x^2 + J_y^2 +J_z^2= const.$ lights up in Fig.~\ref{Fig:SmTcut1chi} (a-c). This can be understood, as the high temperature expansion of $C_m$ only depends on the squared sum of spin XXZ interactions. As $T_{\rm cut}$ further moves to lower temperatures, the oval ring gradually breaks and eventually converges to two [Fig.~\ref{Fig:SmTcut1chi}(d-f)] or one [Fig.~\ref{Fig:SmTcut2chi}(d-f)] bright points, depending on whether $\chi_{xy}$ data are included or not. From these panels, we also see that the fittings, although using only small-size ED results, are rather robust as the parameter points found are rather stable as $T_{\rm cut}$ moves to lower temperatures.

\section{TMGO fitting results}
\label{SMSec:TMGO}

\begin{figure}[htbp]
\includegraphics[angle=0,width=1\linewidth]{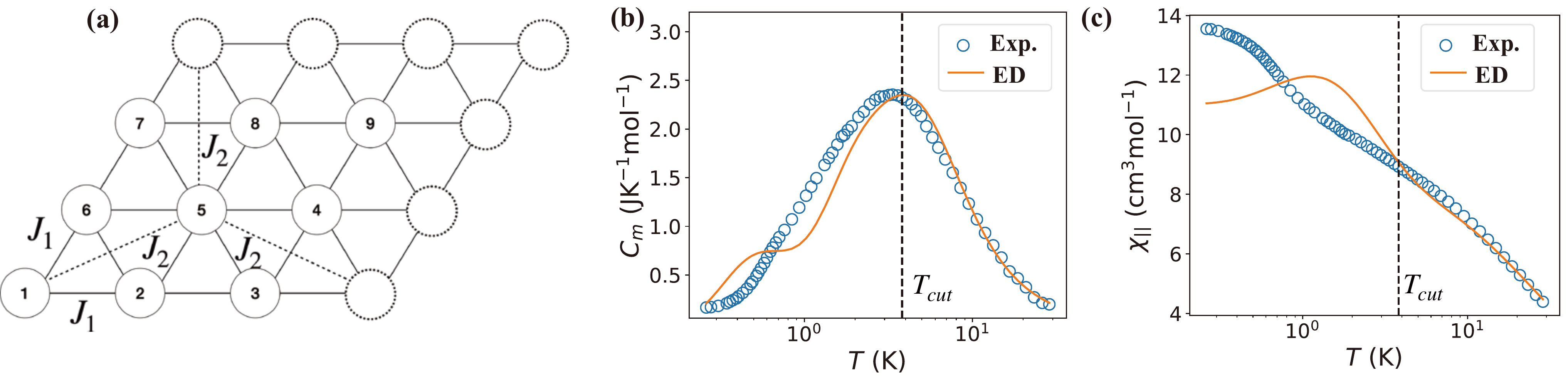}
\caption{(a) The 9-site triangular lattice with both NN and NNN interactions, and PBC on both directions. (b) Experiment data of the TMGO specific heat  and ED fitting with optimal parameter found ($J_1=11.57$ K $J_2=0.89$ K, $\Delta=5.32$ K, and $g=13.64$). (c) Experiment data of susceptibility and ED fitting with the same parameter.}

\label{Fig:SmSecTMGO}
\end{figure}

A 9-site PBC ED calculation is used for fitting [c.f. Fig.~\ref{Fig:SmSecTMGO}] (a), and $T_{cut}$ is set to the peak of specific heat $C_m$. 
Both $C_m$ and $\chi_{||}$ are used for fitting, and the ED results with the optimal parameter 
are shown in Fig.~\ref{Fig:SmSecTMGO} (b, c).

\end{document}